\documentclass[a4paper,twocolumn,english,aps,prl,superscriptaddress]{revtex4-1}
\usepackage[T1]{fontenc}
\usepackage[latin9]{inputenc}
\usepackage{xcolor}
\usepackage{babel}
\usepackage{verbatim}
\usepackage{float}
\usepackage{booktabs}
\usepackage{bm}
\usepackage{amsmath}
\usepackage{amssymb}
\usepackage{graphicx}
\usepackage[pdfusetitle,
 bookmarks=true,bookmarksnumbered=false,bookmarksopen=false,
 breaklinks=false,pdfborder={0 0 1},backref=false,colorlinks=false]
 {hyperref}

\makeatletter

\usepackage{booktabs} 
\usepackage{array}    
\usepackage{amsmath}  

\usepackage{graphicx} 
\usepackage{makecell} 
\usepackage{babel}

\makeatother

\begin{document}
\title{Random-Flux-Induced Transition Sequence between Weak and Strong Topological
Phases with Anisotropic Localization Properties}
\author{Chang-An Li}
\email{changan.li@uni-wuerzburg.de}

\affiliation{Institute for Theoretical Physics and Astrophysics, University of
Würzburg, 97074 Würzburg, Germany}
\author{Bo Fu}
\email{fubo@gbu.edu.cn}

\affiliation{School of Sciences, Great Bay University, Dongguan 523000, Guangdong,
China}
\author{Jian Li}
\affiliation{Department of Physics, School of Science, Westlake University, Hangzhou
310024, Zhejiang, China}
\author{Björn Trauzettel}
\email{trauzettel@physik.uni-wuerzburg.de}

\affiliation{Institute for Theoretical Physics and Astrophysics, University of
Würzburg, 97074 Würzburg, Germany}
\affiliation{Würzburg-Dresden Cluster of Excellence ct.qmat, Germany}
\begin{abstract}
We demonstrate that random flux is able to drive nontrivial topological
phase transitions, in particular between weak topological insulators
(WTIs) and Chern insulators (CIs), illustrated on an anisotropic Wilson-Dirac
model in two dimensions. Remarkably, an intriguing topological transition
sequence WTIs$\rightarrow$CIs$\rightarrow$WTIs occurs with the reentrance
to a WTI but of different weak topology, which is unattainable
with chemical potential disorder. The involvement of anisotropy
and weak topology in such a transition gives rise to emergent quasi-critical
points, where eigen states are extended in one spatial direction but
localized in the other one. This new quantum criticality lies outside the conventional quantum Hall universality class. We provide a comprehensive characterization
of the random-flux-induced phase transitions and quantum criticality
from both bulk and boundary perspectives. Our results describe a qualitatively
new disorder effect based on the interplay of random flux with topological
phases of matter.
\end{abstract}
\date{\today}

\maketitle
\section{Introduction}
Disorder plays
a pivotal role in various physical phenomena, such as Anderson localizations
\cite{Anderson58pr,Evers08rmp} and quantum transport \cite{PALee85rmp,Nazarovbook}.
In the realm of topological phases of matter \cite{Kane10rmp,QiXL11rmp},
the interplay between disorder and topology gives rise to novel topological
phases  \cite{LiJian09prl,Groth09prl,JiangH09prb,Prodan11prb,Yamakage13prb,Kobayashi13prl,Shem14prl,Titum15prl,Pixely15prl,LiC17prl,Stutzer18nature,Meier19Science,ChenR19prb,LiCA19prb,LiCA20prl,LiuG20prl,ZhangDW20scp,Nakajima21NP,SongZD21prl,CuiX22prl,Lapierre22prl,ChenX23prb} and intriguing quantum critical phenomena \cite{WeiHP88prl,Huckestein95RMP,Yoichi02prl,Onoda07prl,Sbierski20prx}.
One of the prototypical examples is the disordered Chern insulator
\cite{Onoda03prl,Thonhauser08prb,Prodan10prl,Prodan11JPA,XueY13prb,Yushihito19prb,Moreno-Gonzalez23AP,Andrews24prb}.
Disorder drives topological phase transitions between normal insulators
and Chern insulators, where the nonzero Chern number leads to extended
bulk states at a critical phase point. Most studies on the impact
of disorder in topological systems so far have focused on on-site
chemical potential disorder. Random flux, genuinely a form of random
magnetic field, whereas represents a distinct type of disorder. It
has been extensively studied in the context of the fractional quantum
Hall effect \cite{Halperin93prb}, high $T_{c}$ superconductivity
\cite{WenXG96prl}, and electron localization properties \cite{Sugiyama93prl,Avishai93prb,Aronov94prb,Sheng95prl,XieXC98prl,Furusaki99prl,Altland99npb,Taras00prl,Cerovski01prb,Markos07prb,Major17pra,LiCA22prb,Wux22prr,Mizoguchi23prb,WangF24nc,QuT24prl},
but its influence on topological phases of matter remains largely
unexplored.

\begin{figure}
\includegraphics[width=1\linewidth]{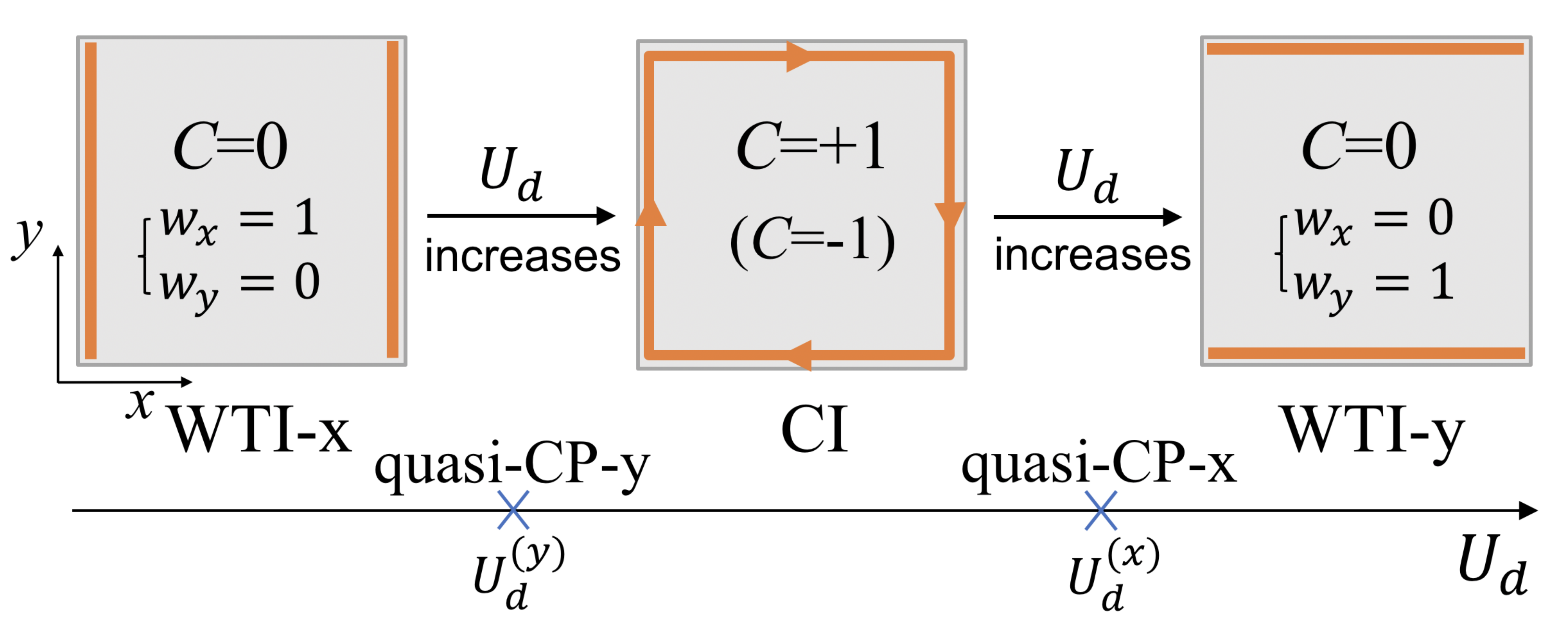}

\caption{Sketch of random-flux-induced phase transitions between weak topological
insulators (WTIs) and Chern insulators (CIs): the transition sequence
WTI-x$\rightarrow$CIs$\rightarrow$WTI-y with increasing random flux
strength $U_{d}$. The topological invariant is labeled as $(C;w_{x}w_{y})$
with $C$ the Chern number and $w_{x/y}$ the weak topological index
along $x/y$ direction, respectively. The evolution of edge state
configurations is indicated by orange lines. During the phase transitions,
there appear quasi-critical points (quasi-CP-x/y) where eigen states
show anisotropic localization properties. \protect\label{fig:Results}}
\end{figure}

Compared with chemical potential disorder, random flux affects physical
properties of a system in fundamentally different ways. First, it
enters the Hamiltonian by coupling to momentum vectors via the Peierls
substitution, modifying the system parameters in a momentum-dependent
manner. Second, it introduces a random $U(1)$ phase factor to the
wavefunctions of charged particles, affecting their coherence and
interference properties. Third, it exhibits diamagnetic effects due
to field fluctuations even at zero mean magnetic field. As such, an
intriguing question is whether random flux can play a fundamentally
new role in the field of topological phases of matter going beyond
the influence of chemical potential disorder.

In this work, we discover the random-flux-induced topological phase
transitions between WTIs and CIs, implying a new type of quantum criticality.
We show that random flux can drive a topological phase transition
sequence WTI-x$\rightarrow$CIs$\rightarrow$WTI-y with the reentrance
to a WTI but the nontrivial direction rotated by $\pi/2$ (see Fig.\ \ref{fig:Results}).
This cannot happen due to chemical potential disorder. We explicitly
demonstrate such a phase transition sequence on an anisotropic Wilson-Dirac
(AWD) model (see Fig.\ \ref{fig:phase1}). These phase transitions
are characterized by quasi-critical points, at which the eigen states
are delocalized in one spatial direction but remain localized in the
other one. We characterize the phase transitions and the consequent
localization properties in terms of real-space topological invariants,
finite-size scaling, and a new analytical theory. 

The article is organized as follows. In Sec. II, we analyse the properties of the anisotropic Wilson-Dirac model and introduce random flux to the 2D lattice. In Sec. III, we present the phase transition sequence between WTIs and CIs driven by random flux. In Sec. IV, we show the corresponding transport features during the phase transitions. In Sec. V, we discuss the phase digram in energy-flux space. In Sec. VI, we explain the underlying mechanism of the phase transition induced by random flux. In Sec. VII, we discuss the emergent quantum criticality at the transition between WTIs and CIs. In Sec. VIII, we propose an analytical theory for the phase transition from a diamagnetic treatment of multiple scattering processes. In Sec. IX, we consider the experimental relevance of our predictions. Finally, we conclude our results with a discussion in Sec. X.

\section{Anisotropic Wilson-Dirac model with random flux}
As a concrete implementation to realize CIs and
WTIs on a 2D lattice, we consider the AWD tight-binding Hamiltonian

\begin{align}
H_{0} & =\sum_{{\bf r}}\sum_{\alpha=x,y}\left[\frac{b_{\alpha}}{2}c_{{\bf r}+{\bf e}_{\alpha}}^{\dagger}\sigma_{z}c_{{\bf r}}+H.c.\right]+\sum_{{\bf r}}mc_{{\bf r}}^{\dagger}\sigma_{z}c_{{\bf r}}\nonumber \\
 & +\sum_{{\bf r}}\sum_{\alpha=x,y}\left[\frac{iv_{\alpha}}{2}c_{{\bf r}+{\bf e}_{\alpha}}^{\dagger}\sigma_{\alpha}c_{{\bf r}}+H.c.\right],\label{eq:Model1}
\end{align}
where $c_{{\bf r}}^{\dagger}$ and $c_{{\bf r}}$ are creation and
annihilation operators on a site ${\bf r}=(x,y)$, ${\bf e}_{x,y}$
is the unit vector along $x(y)$ direction. $\sigma_{x,y,z}$ are
Pauli matrices for orbital degrees of freedom. Here, $m$ (mass term),
$b_{\alpha}$, and $v_{\alpha}$ are model parameters. The anisotropy
of this model arises from the choice $b_{x}\neq b_{y}$ and $v_{x}\neq v_{y}$.
It is a generalization of the CI model \cite{Qi06prb,Bernevig06Scien}.

The Bloch
Hamiltonian for the AWD model reads
\begin{equation}
H_{0}({\bf k})={\bf d}({\bf k})\cdot\bm{\sigma},\label{eq:Model}
\end{equation}
where
\begin{equation}
{\bf d}({\bf k})\equiv(v_{x}\sin k_{x},v_{y}\sin k_{y},m+\sum_{\alpha=x,y}b_{i}\cos k_{i}),
\end{equation}
and $\bm{\sigma}=(\sigma_{x},\sigma_{y},\sigma_{z})$ represents the
Pauli matrices. The vector ${\bf k}\equiv(k_{x},k_{y})$ is the Bloch
wave vector. Here, $m$ is the mass term, and $b_{i}$ and $v_{i}$
are model parameters.
The Chern number at half-filling can be calculated as

\begin{align}
C= & \int_{BZ}\frac{d{\bf k}}{4\pi|{\bf d}({\bf k})|^{3}}{\bf d}({\bf k})\cdot\partial_{k_{x}}{\bf d}({\bf k})\times\partial_{k_{y}}{\bf d}({\bf k}).
\end{align}
For simplicity, we define $b_{\pm}\equiv|b_{x}\pm b_{y}|$ and assume
$b_{+}>b_{-}$ and $v_{x}v_{y}>0$. The Chern number takes 
\begin{align}
C(m)= & \begin{cases}
0, & |m|>b_{+};\\
+1, & -b_{+}<m<-b_{-};\\
0, & |m|<b_{-};\\
-1, & b_{-}<m<b_{+}.
\end{cases}\label{eq:Chernnumber}
\end{align}

\begin{figure}
\includegraphics[width=1\linewidth]{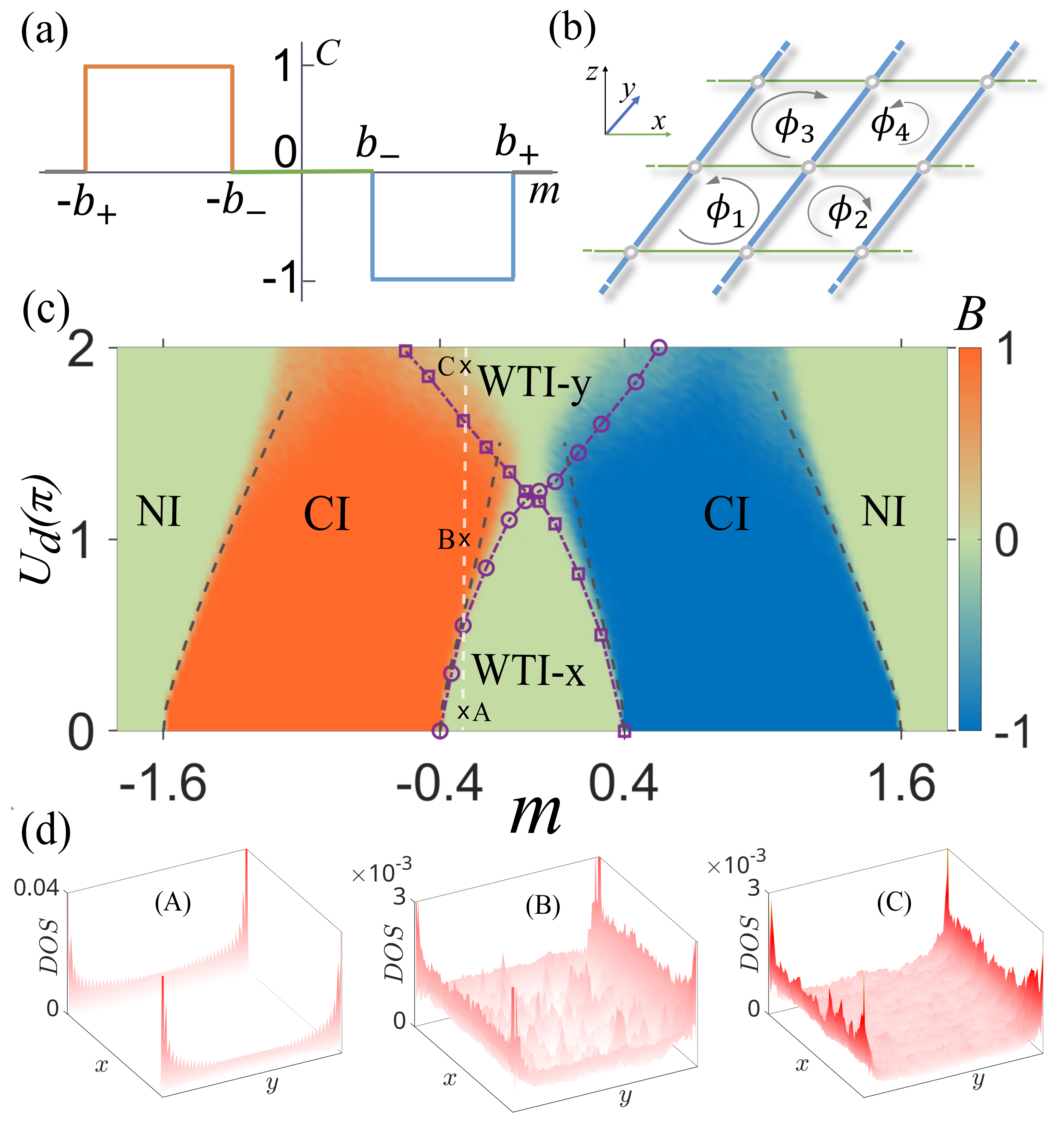}

\caption{(a) Chern number as a function of mass $m$ for the AWD model in the
clean limit. We define $b_{\pm}\equiv|b_{x}\pm b_{y}|$, with the
assumptions $b_{+}>b_{-}$ and $v_{x}v_{y}>0$. (b) Schematic of adding
random flux to the AWD model on a square lattice. Here, $\phi_{i=\{1,2,3,4\}}$
exemplify random fluxes. (c) Phase diagram of the AWD model in the
presence of random flux. It illustrates the Bott index (Chern number)
as a function of random flux strength $U_{d}$ and mass $m$ at half-filling.
The dashed purple lines indicate the phase boundaries obtained from
the effective Hamiltonian on configuration average, where the circle
and square represent gap-closing at HSPs $X$ and $Y$, respectively.
The dashed black lines indicates phase boundaries obtained from the
analytical method. Other parameters are: $b_{x}=1$, $b_{y}=0.6$,
$v_{x}=0.2$, and $v_{y}=1$. The system size is $L_{x}\times L_{y}=30\times30$
with periodic boundary conditions. We average over $120$ random flux
configurations. (d) The evolvement of edge state configurations corresponding
to phase points A, B, and C in (c) for $m=-0.3$ with $U_{d}=0.1\pi,1.0\pi,1.9\pi$,
respectively. \protect\label{fig:phase1}}
\end{figure}

The phase diagram in terms of Chern numbers is shown in Fig.\ \ref{fig:phase1}(a).
Notably, an additional phase with $C=0$ emerges for $|m|<|b_{x}-b_{y}|$
if $b_{x}\neq b_{y}$. In this phase, a pair of gapless edge states
appears at one boundary but not at the other boundary (see Appendix A).
It is characteristic of a WTI phase, where the topological protection
and boundary states depend on specific crystal directions \cite{Yoshimura14prb,FuL07prb,Hughes11prb}.
The Chern number $C=0$, however, does not distinguish between normal
insulators (NIs) and WTIs.
To address this, we introduce two weak $Z_{2}$ indices, $(w_{x},w_{y})$,
based on parity configurations at HSPs. These indices complement the
strong index $C=0$ and identify the WTI. We note that the AWD model respects an inversion symmetry

\begin{equation}
\mathcal{P}H_{0}({\bf k})\mathcal{P}^{-1}=H_{0}(-{\bf k}),
\end{equation}
where $\mathcal{P}=\sigma_{z}$. In this case, the topology of the
system can be fully determined by examining four high-symmetry points
(HSPs). The Chern number is directly connected to the parity of occupied
bands at these HSPs through the relation \citep{Hughes11prb}
\begin{equation}
(-1)^{C}=\Pi_{j=1}^{4}\xi_{j}(K_{j}),
\end{equation}
where $K_{j}\in\{K_{1}=\Gamma(0,0),K_{2}=X(\pi,0),K_{3}=M(\pi,\pi),K_{4}=Y(0,\pi)\}$,
and $\xi_{j}=\pm1$ is the parity value defined by $\mathcal{P}\psi_{-}(K_{j})=\xi_{j}\psi_{-}(K_{j}).$
At these HSPs, the parity is related to the effective mass by

\begin{equation}
\xi_{j}=-\mathrm{sgn}(\tilde{m}_{K_{j}}).
\end{equation}
The gap-closing transition flips the sign of the effective mass term
at HSPs, which in turn changes the parity values and thus changes the Chern number.

In a WTI, the parity configuration
$\xi_{j}$ at HSPs takes two positive and two negative values, while
for a NI, parity values $\xi_{j}$ at four HSPs are all positive or
all negative. 
To illustrate this, let us reduce the dimensionality to 1D by setting
$k_{y}^{*}=0$ or $\pi$, leading to the following Hamiltonian

\begin{equation}
H(k_{x},k_{y}^{*}=0/\pi)=(m\pm b_{y}+b_{x}\cos k_{x})\sigma_{z}+v_{x}\sin(k_{x})\sigma_{x}.
\end{equation}
By applying a unitary transformation with $U(\theta)=\exp(-i\frac{\pi}{4}\sigma_{z})\exp(-i\frac{\pi}{4}\sigma_{x})$,
the Hamiltonian transforms to 

\begin{equation}
\tilde{H}(k_{x},k_{y}^{*})=(m\pm b_{y}+b_{x}\cos k_{x})\sigma_{x}+v_{x}\sin(k_{x})\sigma_{y}.
\end{equation}
This 1D Hamiltonian respects inversion symmetry as well as chiral
symmetry. Its topology is characterized by a quantized polarization
$p_{x}=0/\frac{1}{2}$ (or equivalently winding number $0/1$). With
the help of inversion symmetry, we obtain

\begin{alignat}{1}
\xi_{1}\xi_{2} & =e^{i2\pi p_{x}(k_{y}^{*}=0)},\\
\xi_{3}\xi_{4} & =e^{i2\pi p_{x}(k_{y}^{*}=\pi)}.
\end{alignat}
Due to the constraint $\xi_{1}\xi_{2}=\xi_{3}\xi_{4}$, we can define
the weak index $w_{x}$ as 
\begin{equation}
(-1)^{w_{x}}=\xi_{1}\xi_{2}.
\end{equation}
To have $w_{x}=1$, it requires 
\begin{equation}
|m+b_{y}|<b_{x}\cap|m-b_{y}|<b_{x}.\label{eq:Cond1}
\end{equation}
For our parameter setting, it gives $|m|<b_{-}$,
which is consistent with the phase diagram. Similarly, we can define the weak index
$w_{y}$ from
\begin{equation}
(-1)^{w_{y}}=\xi_{1}\xi_{4}.
\end{equation}
Therefore, the AWD model
thereby distinguishes three distinct phases for $C=0$ by $(C;w_{x}w_{y})=(0;00),(0;10)$,
and $(0;01)$ corresponding to NIs, WTI-x, and WTI-y, respectively.

We apply random flux to the AWD model in 2D real space. As illustrated
in Fig.\ \ref{fig:phase1}(b), a magnetic flux with random value
$\phi({\bf r})$ is enclosed within each plaquette of the square lattice.
The random value $\phi({\bf r})$ is uniformly distributed within
$[-U_{d}/2,U_{d}/2]$ with $U_{d}$ the random flux strength \cite{Sheng95prl,XieXC98prl,Furusaki99prl,LiCA22prb}.
The random flux is connected to a random magnetic field as $\phi({\bf r})=B({\bf r})$.
It thus affects the Hamiltonian through a vector potential ${\bf A}({\bf r})$
via the Peierls substitution. Generally, the hopping terms are modified as $t_{\langle {\bf r}_m{\bf r}_n\rangle}\rightarrow t_{\langle {\bf r}_m{\bf r}_n\rangle}\exp[i\theta_{{\bf r}_m{\bf r}_n}]$, where the $U(1)$ phase factors are $\theta_{{\bf r}_m{\bf r}_n}=\int_{{\bf r}_m}^{{\bf r}_n}{\bf A}({\bf r})\cdot d{\bf r}$ and $t_{\langle {\bf r}_m{\bf r}_n\rangle}$ indicates the nearest-neighbor hopping between sites ${\bf r}_m$ and ${\bf r}_n$. Note that different gauge choice does not change the results. In the following,
we take the gauge choice ${\bf A}({\bf r})=(\theta({\bf r}),0,0)$
where $\theta(x,y+1)-\theta({\bf r})=\phi({\bf r})$ \cite{Furusaki99prl,LiCA22prb}.
The hopping terms in Eq.\ \eqref{eq:Model1} along $x$ direction
are modified to $\sum_{{\bf r}}\frac{1}{2}\exp[i\theta({\bf r})]c_{{\bf r}+{\bf e}_{x}}^{\dagger}(b_{x}\sigma_{z}+iv_{x}\sigma_{x})c_{{\bf r}}+H.c.$.
Upon introducing random flux into the system, we employ the Bott index
$B$ to characterize the bulk topology, which has been proven to be
equivalent to the Chern number \cite{Toniolo22LMP}. The Bott index
is defined as $B=\frac{1}{2\pi}\mathrm{Im\ Tr[log}(\tilde{U}_{y}\tilde{U}_{x}\tilde{U}_{y}^{\dagger}\tilde{U}_{x}^{\dagger})]$
\cite{Loring10epl}, where $\tilde{U}_{x}$ and $\tilde{U}_{y}$ are
the reduced matrices of $U_{x}=Pe^{i2\pi\hat{x}/L_{x}}P$ and $U_{y}=Pe^{i2\pi\hat{y}/L_{x}}P$
in the occupied space, respectively. In the
above formula, $\hat{x}(\hat{y})$ is the position operator along
the $x(y)$ dimension and $L_{x}$($L_{y}$) is the corresponding
size, and $P$ is the projection operator.

\section{Phase transition sequence between weak topological insulators and Chern insulators}
The phase diagram modified by the presence of random
flux is depicted in Fig.\ \ref{fig:phase1}(c). Due to its ``symmetric''
with respect to $m=0$, we focus on the region $m<0$. Let us start
with WTIs. For the chosen parameters, the WTI-x has $(C;w_{x}w_{y})=(0;10)$
in the clean limit. It becomes apparent if we follow the line $m=-0.3$ [Fig.\ \ref{fig:phase1}(c) and Fig.\ \ref{fig:transport}(c)].
As $U_{d}$ increases, a topological phase transition sequence occurs.
The WTI-x is first driven to an insulating phase with $C=+1$. Note
that this random-flux-induced topological nontrivial phase is equivalent
to a CI \cite{Prodan11prb}. Remarkably, increasing $U_{d}$ further,
the system reenters to a WTI but with the nontrivial direction rotated
by $\pi/2$, i.e., the WTI-y phase with $(C;w_{x}w_{y})=(0;01)$.
Hence, we discover a topological phase transition sequence WTI-x$\rightarrow$CIs$\rightarrow$WTI-y
sketched in Fig.\ \ref{fig:Results}. It is directly confirmed by
the redistribution of edge states shown in Fig.\ \ref{fig:phase1}(d)
and further verified by quantum transport signatures discussed below.
Let us emphasize that tuning the mass term $m$ only drives phase
transitions between CIs, WTI-x, and NIs in the clean limit {[}Fig.\ \ref{fig:phase1}(a){]}.
Therefore, the emergent WTI-y phase is a distinct phase that indicates
modifications of other model parameters in Eq.\ \eqref{eq:Model1}
such as $b_{\alpha}$, which cannot be achieved by chemical potential
disorder. Indeed, on-site chemical potential disorder is incapable
of driving phase transitions between WTIs and CIs (see Appendix C).

Notably, the fate of CIs under random flux with different $m$ values
can be very different. The CIs can be driven to NIs at large $U_{d}$
{[}e.g., for $m=-1.5$ in Fig.\ \ref{fig:transport}(a){]}. Moreover,
we find that the CI can also be driven to the WTI-y directly (e.g.,
for $m=-0.42$ in Fig.\ \ref{fig:transport}(b)) or keeps as CI as $U_{d}$ grows [e.g., for $m=-0.8$ Fig.\ \ref{fig:transport}(b)].
In addition, the topological Anderson transition from NIs to CIs is
not driven by random flux. All of these features suggest different
mechanisms of topological phase transitions induced by random flux,
which we specify in the following.

\section{Transport feature during the phase transition sequence}
\begin{figure}
\textcolor{blue}{\includegraphics[width=1\linewidth]{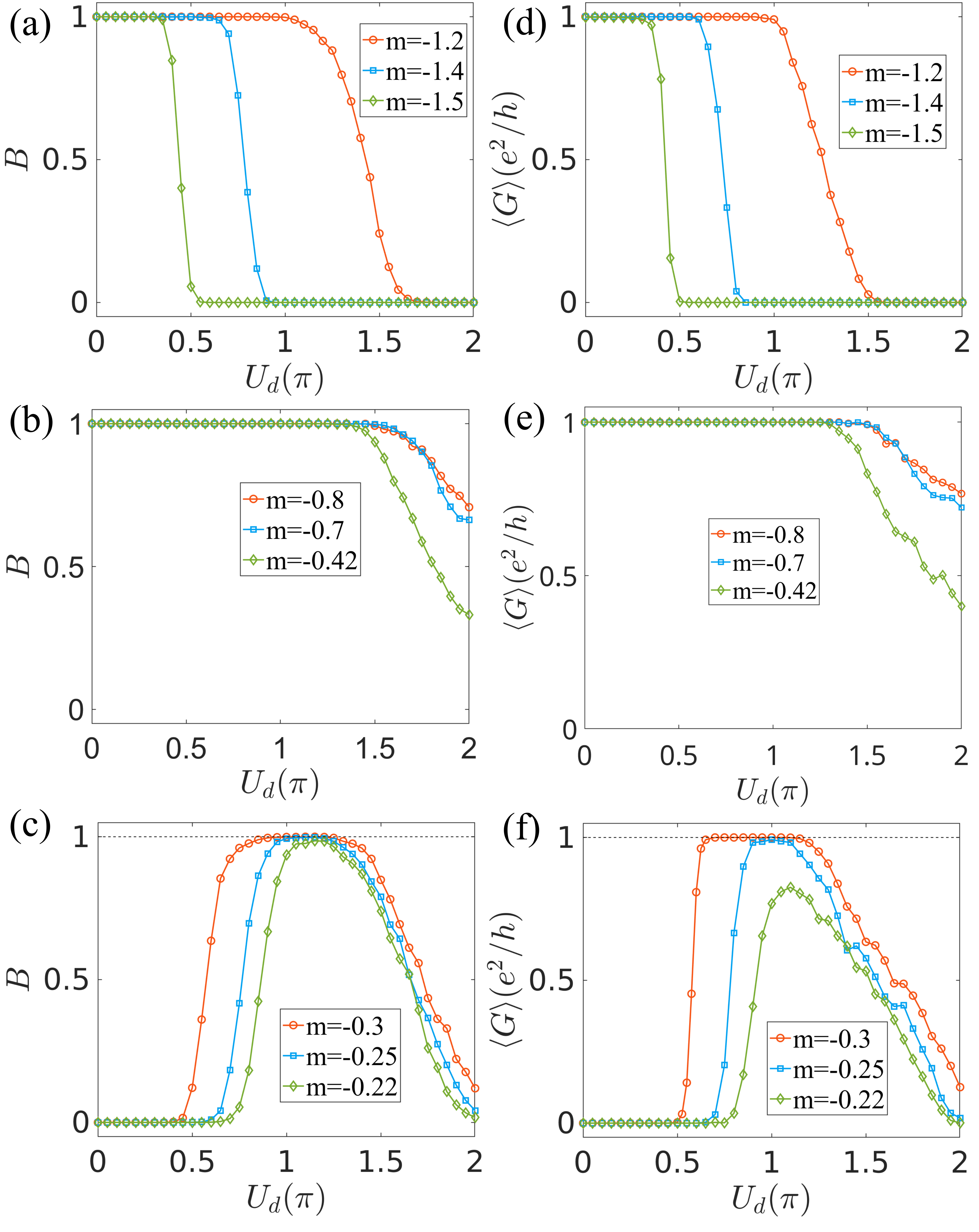}}
\caption{(a-c) Bott index as a function of $U_{d}$ for different mass
$m$, averaging over $1280$ random flux configurations for $L_{x}\times L_{y}=30\times30$. 
(d-f): Averaged two-terminal conductance along $x$ direction
as a function of $U_{d}$ corresponding to (a-c), respectively.
Here, the system size is $L_{x}\times L_{y}=200\times200$, the Fermi
energy is $E=0$, and we average over $200$ random flux configurations.
\label{fig:transport}}
\end{figure}

We further verify the phase transitions more closely by examining the change of
Bott index and corresponding transport signatures. 
The chiral edge modes corresponding to nonzero Chern
numbers give rise to quantized conductance. To this end, we calculate
the two-terminal conductance using the Landauer-B$\mathrm{\ddot{u}}$ttiker
formalism. The conductance $G$ can be evaluated as 
\begin{equation}
\ensuremath{G(E)=\frac{e^{2}}{h}\mathrm{Tr}[\Gamma_{L}G^{r}\Gamma_{R}G^{a}]},
\end{equation}
where $\ensuremath{G^{r,a}}$ are the retarded and advanced Green's
functions, respectively, and $\Gamma_{L,R}$ are the line-width functions
coupling two terminals to the central region of interest. Here, $e$
is the electron charge and $h$ is the Planck constant. Upon introducing
random flux, the conductance $G=\frac{e^{2}}{h}$ survives weak random
flux for CIs until a topological phase transition occurs, as shown
in Fig.\ \ref{fig:transport}(d). In Fig.\ \ref{fig:transport}(e),
conductance drops to some extent at large $U_{d}$ corresponding to
Fig.\ \ref{fig:transport}(b). Starting from the WTI, the conductance
increases from zero to a plateau $G=\frac{e^{2}}{h}$ as increasing
$U_{d}$ and decreases gradually close to zero {[}Fig.\ \ref{fig:transport}(f){]},
consistent with the topological phase transition sequence observed
in Fig.\ \ref{fig:transport}(c). The emergent conductance plateau
signals the formation of chiral edge modes directly.

\section{Phase diagram in energy-flux space}
We present the phase diagram of the AWD model under random flux in the parameter
space $(E,U_{d})$. To this end, we employ the the non-conmmutative Kubo formula for the calculation
of Chern number \citep{Prodan10prl,Prodan11JPA}. It is calculated as 
\begin{alignat}{1}
C & (E)=\frac{-2\pi i}{L^{2}}\mathrm{Tr}\left(P\left[-i[\hat{x}_{1},P],-i[\hat{x}_{2},P]\right]\right),
\end{alignat}
where the $P$ is the projection operator of occupied states on the
basis of real space $L\times L$ lattice. In the above formula, the
``commutator'' is given by
\begin{alignat}{1}
-i[\hat{x_{i}},P] & =\sum_{m=1}^{q}c_{m}\left(e^{-im\Delta\hat{x}_{i}}Pe^{+im\Delta\hat{x}_{i}}-e^{+im\Delta\hat{x}_{i}}Pe^{-im\Delta\hat{x}_{i}}\right),
\end{alignat}
where $\Delta\equiv\frac{2\pi}{L}$, $\hat{x}_{1,2}\equiv\hat{x},\hat{y}$,
and $q$ is an integer $q\leq L/2$. With a larger $q$, the results
get more accurate but it becomes more time consuming. Here, $c_{m}$
is the coefficient determined by the solution of the linear equation
\begin{alignat}{1}
\hat{M}\left(\begin{array}{c}
c_{1}\\
c_{2}\\
c_{3}\\
\vdots\\
c_{q}
\end{array}\right) & =\frac{1}{2\Delta}\left(\begin{array}{c}
1\\
0\\
0\\
\vdots\\
0
\end{array}\right),
\end{alignat}
where $\hat{M}_{ij}=j^{2i-1},\ i,j=1,2,3,...,q.$

\begin{figure}
\includegraphics[width=1\linewidth]{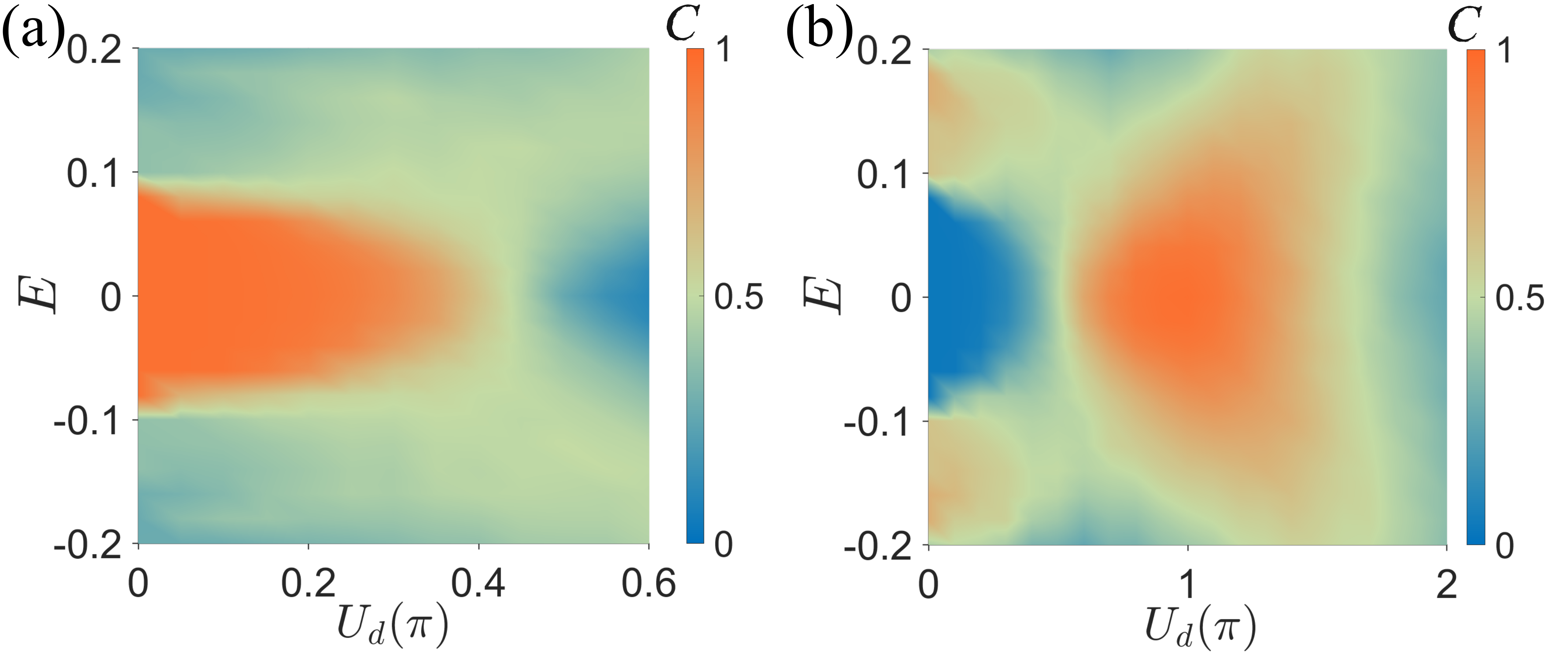}

\caption{The calculated Chern number in the space $(E,U_{d})$ for different
parameter $m$. (a) $m=-1.5$ and (b)  $m=-0.3$. Other parameters are:
$b_{x}=1,b_{y}=0.6,v_{x}=0.2$, $v_{y}=1,$ $L=40,$ and $q=12$.}
\label{fig:Phase_prodan}
\end{figure}

We plot the disorder averaged Chern number in the
space $(E,U_{d})$ for different parameter $m$ in Fig. \ref{fig:Phase_prodan}. For CIs with $m=-1.5$  at
$U_{d}=0$,
the Chern number is $C=+1$ within the band gap $|E_{g}|<0.1$ [see Fig. \ref{fig:Phase_prodan}(a)]. As increasing $U_{d}$, the Chern number drops close to
zero, entering an NI. Different from the on-site disordered
CIs \citep{Prodan10prl,Prodan11JPA}, the phase boundary
narrows down monotonically in energy as increasing $U_{d}$
and no topological Anderson transition happens. 

Then let us turn to the WTI-x with $m=-0.3$ {[}Fig.
\ref{fig:Phase_prodan}(b){]}. At $U_{d}=0$, we find that the Chern
number is zero within the band gap $|E_{g}|<0.1$. As increasing the
random flux strength $U_{d}$, the Chern number jumps to $C=+1$,
entering a CI. The Chern number keeps at $C\simeq+1$
in a relatively large energy window. With $U_{d}$ growing further,
the Chern number drops to a value close to zero again. This result
is consistent with the topological phase transition sequence WTI-x$\rightarrow$CIs$\rightarrow$WTI-y
we have discussed. Therefore, the transition sequence exists in a
finite energy window.

\section{Momentum-dependent renormalizations of effective mass and parity flips at high-symmetry points (HSPs)}
To uncover the underlying physical mechanism, we first derive an effective
Hamiltonian by averaging over random flux configurations. We average
a large number ($\sim4\times10^{5}$) of random flux configurations
such that the translation and inversion symmetries are effectively
restored. The averaged Green's function is expressed as $G_{\mathrm{avg}}^{r}({\bf r}-{\bf r}',E)=\langle G^{r}({\bf r},{\bf r}',E)\rangle$,
where $\langle...\rangle$ denotes the average over random flux configurations.
Performing a Fourier transformation on the averaged Green's function
$G^{r}({\bf k},E)=\sum_{{\bf r}}G_{\mathrm{avg}}^{r}({\bf r},E)e^{i{\bf k}\cdot{\bf r}}$,
we construct an effective Hamiltonian $H_{\mathrm{eff},G}({\bf k})=-[G^{r}({\bf k},E=0)]^{-1}$ (see Appendix D). It enables us to extract the effective mass terms
and determine the parity configurations at HSPs during phase transitions.

\begin{figure}
\includegraphics[width=1\linewidth]{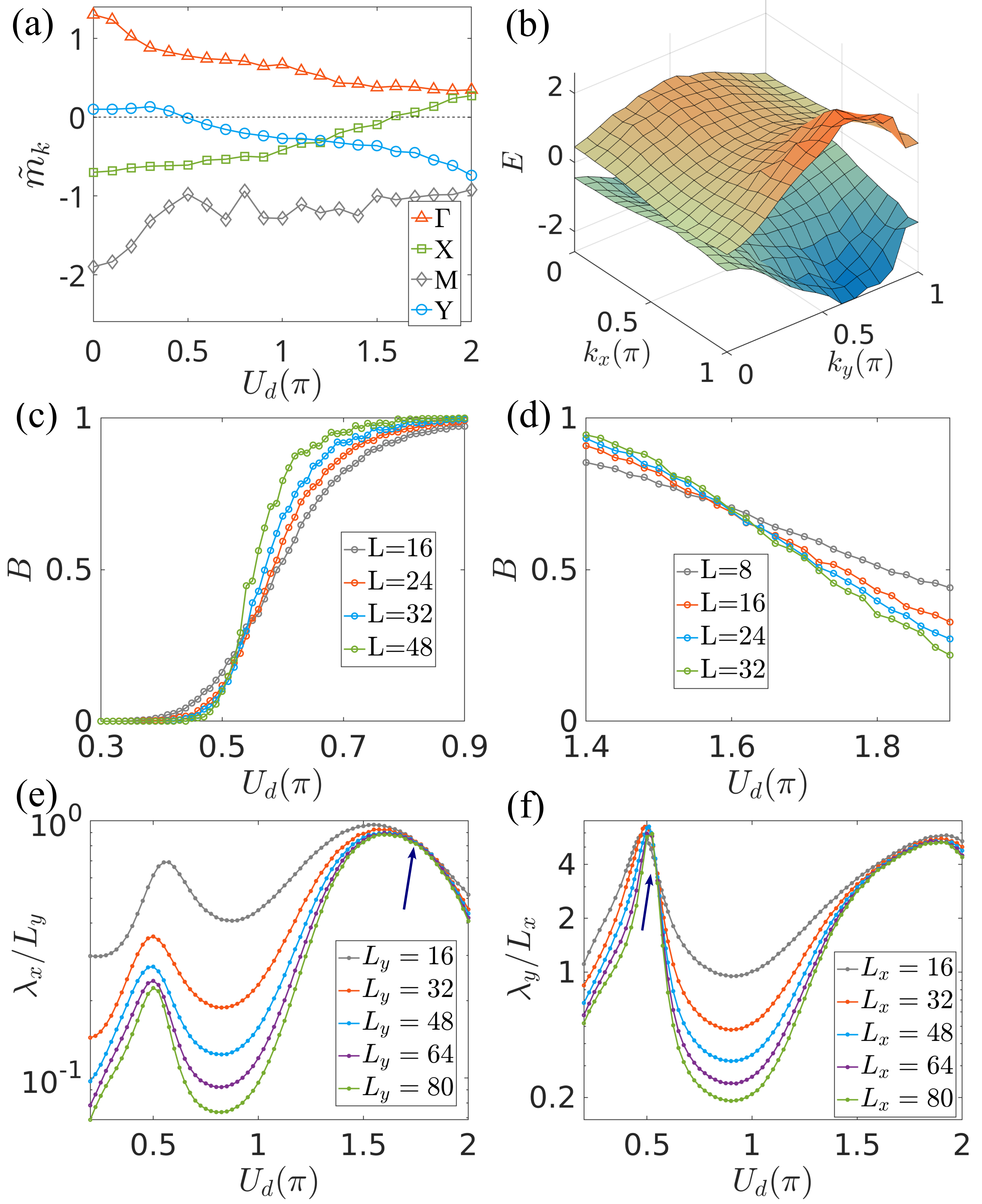}

\caption{(a) Renormalized mass at different HSPs obtained from the effective
Hamiltonian $H_{\mathrm{eff},G}({\bf k})$ with $m=-0.3$. Here, $4\times10^{5}$
random flux configurations are averaged at each point. (b) Band structure
at $m=-1.2$ obtained from the effective Hamiltonian. (c) and (d):
Scaling of Bott index as a function of $U_{d}$. (e) and (f): Localization
length $\lambda_{x}$ and $\lambda_{y}$ as a function of $U_{d}$
on a quasi-1D tube of length $2\times10^{6}$, where $L_{x/y}$ indicates
the width (circumference). Other parameters are: $m=-0.3,b_{x}=1$,
$b_{y}=0.6$, $v_{x}=0.2$, and $v_{y}=1$. \protect\label{fig:mass}}
\end{figure}

\begin{table}[H]
\centering

\caption{Evolvement of topological invariant, parity indicator, and edge state
configurations during the topological transition sequence WTI-x$\rightarrow$CIs$\rightarrow$WTI-y
induced by random flux.}

\label{table:parameters}

\begin{tabular}{cccc}
\hline
 \hline
Transition sequence: & \multicolumn{3}{c}{WTI-x\enskip{}$\longrightarrow$\enskip{}CIs\quad{}$\longrightarrow$\enskip{}WTI-y\quad{}}\\
\hline
Top. inv. $(C;w_{x}w_{y})$: & (0;10) & $C=\pm$1 & (0;01)\\
\hline
Parity indicator $\bm{\xi}$:
& $\left(\begin{array}{cc}
- & +\\
- & +
\end{array}\right)$ & $\left(\begin{array}{cc}
\pm & +\\
- & \text{\ensuremath{\pm}}
\end{array}\right)$ & $\left(\begin{array}{cc}
+ & +\\
- & -
\end{array}\right)$\vspace{0.5mm}\\
\hline
Edge states: & x-boundary & all edges & y-boundary\\
\hline
\hline
\end{tabular}
\end{table}

Specifically, we take the phase transition sequence WTI-x$\rightarrow$CIs$\rightarrow$WTI-y
as an example, depicted in Fig.\ \ref{fig:mass}(a). We focus on
the effective masses and parities at the HSPs $\{\Gamma,X,M,Y\}$.
In the clean limit with $m=-0.3$, the effective masses at HSPs have
$\tilde{m}_{\Gamma,Y}>0$ and $\tilde{m}_{X,M}<0$, corresponding
to a WTI-x with $(C;w_{x}w_{y})=(0;10)$. Random flux affects the
renormalization of effective mass at the four HSPs in different patterns.
At the $Y$ point, $\tilde{m}_{Y}$ decreases to zero at around $U_{d1}\simeq0.5\pi$,
and continues to decrease, signaling a gap-closing transition. As
$U_{d}$ increases further, $\tilde{m}_{X}$ gradually shifts from
negative to positive values, indicating another gap-closing transition
at $X$ near $U_{d2}\simeq1.6\pi$. Between $U_{d1}$ and $U_{d2}$,
the system enters the CI phase with $C=+1$. The band structure obtained
from an effective Hamiltonian at $U_{d}=1.2\pi$ is plotted in Fig.\ \ref{fig:mass}(b).
In contrast, the masses at the other two HSPs remain unchanged with
$\tilde{m}_{\Gamma}>0$ and $\tilde{m}_{M}<0$ throughout the process.
We define the parity indicator at HSPs as $\bm{\xi}\equiv\left(\begin{array}{cc}
\xi_{Y} & \xi_{M}\\
\xi_{\Gamma} & \xi_{X}
\end{array}\right)$. During the topological phase transition sequence WTI-x$\rightarrow$CIs$\rightarrow$WTI-y,
the evolutions of topological invariant, parity indicator, and edge
state configurations are summarized in Table\ \ref{table:parameters}.
These results align with the phase diagram in Fig.\ \ref{fig:phase1}(c).

\section{Emergent new quantum criticality at the transition between weak topological insulators and Chern insulators}
The plateau transition
$C:0\leftrightarrow\pm1$ typically marks a critical point where the
localization length diverges \cite{Onoda03prl}. In contrast, we find
that the weak topology and anisotropy both matter for the transition
sequence WTI-x$\rightarrow$CIs$\rightarrow$WTI-y, giving rise to
a new quantum criticality. The scaling behavior of the Bott index
for this transition sequence is illustrated in Figs.\ \ref{fig:mass}(c)
and \ref{fig:mass}(d). As the system size $L$ increases, the Chern
number exhibits opposite trends on either side of two different phases.
In Figs.\ \ref{fig:mass}(e) and \ref{fig:mass}(f), we plot the
renormalized localization lengths \cite{MacKinnon83ZP,Yamakage13prb}
corresponding the transitions sequence. Notably, we observe pronounced
spatial anisotropy in the localization behavior: At the transition
point $U_{d}^{(y)}$, the localization length $\lambda_{y}$ diverges
in the thermodynamic limit, while $\lambda_{x}$ is finite. This indicates
that the states at this critical point are extended in $y$ direction
but remain localized in $x$ direction. In contrast, at the transition
point $U_{d}^{(x)},$ $\lambda_{x}$ diverges while $\lambda_{y}$
becomes finite. Thus the system exhibits anisotropic localization
properties at different disorder strength. We identify these phase
transition points as ``quasi-critical points'' (quasi-CPs), characterized
by anisotropic localization behavior along two spatial directions.
Near these quasi-critical points, we perform a single-parameter scaling
analysis using a universal function of the form $F(f_{1}(u)L^{1/\nu},f_{2}(u)L^{-\eta})$
with $\nu$ the critical exponent, $u\equiv(U_{d}/U_{d}^{(x,y)}-1)$,
and $\eta$ an auxiliary parameter \cite{LuoX21prl,LiCA22prb}. $f_{1}(u)L^{1/\nu}$
and $f_{2}(u)L^{-\eta}$ are relevant and irrelevant functions, respectively.
By Taylor expansion of the scaling function near quasi-CPs, we obtain
$\nu\simeq5.46\pm0.75$, different from the critical exponent $\nu\simeq2.59$
in quantum Hall transitions \cite{Slevin09prb}. This new critical
phenomenon is interpreted as a consequence of anisotropy and weak
topology of the system in the transition sequence, reflected by topological
invariant change $(C;w_{x}w_{y})=(0;10)\leftrightarrow C=\pm1\leftrightarrow(C;w_{x}w_{y})=(0;01)$.
It has been shown that the weak topology can also alter quantum criticality
in other 2D systems \cite{ZhaoP24prl,LiCA22FP}. We note that such quasi-CPs
are absent in the transition between CIs and NIs (see Appendix E).

\section{Diamagnetic effects and multiple scattering from random flux}
\begin{figure}
\includegraphics[width=1\linewidth]{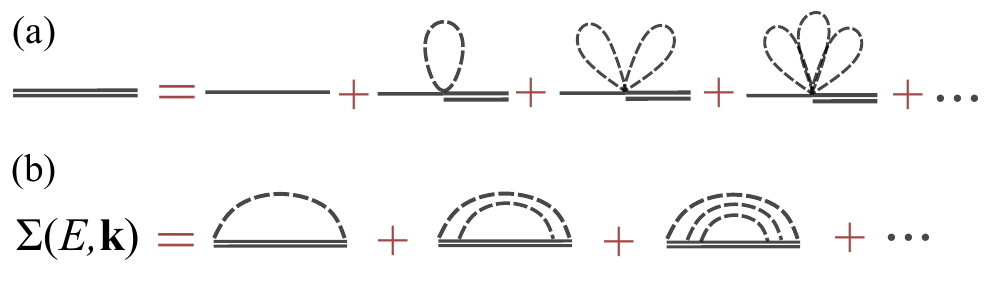}

\caption{(a) Dyson equation of Dirac electrons being scattered by random flux.
The single (double) line indicates the bare (dressed) Green's function
of electrons and dashed lines indicate the scattering by vector potential.
(b) Self-energy calculation using a self-consistent approach under
the non-crossing approximation. \protect\label{fig:self-energy}}
\end{figure}

We develop an analytical understanding
of the random-flux-driven topological phase transition by considering
the diamagnetic effect and the non-crossing scatterings between electrons
and random flux (see Fig.\ \ref{fig:self-energy}). In presence
of random flux, the full Hamiltonian is expressed as $H=H_{0}+V[{\bf A}({\bf r})]$,
where $H_{0}$ is the Hamiltonian in clean limit and $V[{\bf A}({\bf r})]$
is the external potential due to random flux. Note that the variance
of the vector potential in $k$-space is $\langle A_{\alpha}({\bf q})A_{\beta}(-{\bf q})\rangle=D_{\alpha\beta}({\bf q})$
and we define $D_{\alpha\beta}({\bf q})\equiv\frac{U_{d}^{2}}{12}\frac{\delta_{\alpha\beta}-\hat{q}_{\alpha}\hat{q}_{\beta}}{|{\bf q}|^{2}}$
with $\hat{q}_{\alpha}\equiv\frac{{\bf q}_{\alpha}}{|{\bf q}|}$. Notably, $D_{\alpha\beta}({\bf q})$ exhibits strong singularity in the forward direction $({\bf q}=0)$ which leads to infrared divergences in self-energy calculations, necessitating a regulation due to the electron-field coupling (see Appendix B).

The random flux leads to a nonzero diamagnetic term due to spatial
fluctuations of random magnetic fields. After random flux averaging,
only the terms containing even orders of $A_{\alpha}({\bf k})$ survive
with ${\bf k}$ the wave vector {[}see Fig.\ \ref{fig:self-energy}(a){]}.
The self-consistent Dyson equation is given by $G^{r}({\bf k},{\bf k}',E)=\delta_{{\bf k}{\bf k}'}G_{0}^{r}({\bf k},E)+G_{0}^{r}({\bf k},E)\sum_{{\bf k}''}V({\bf k},{\bf k}'')G^{r}({\bf k}'',{\bf k}',E)$,
where $G_{0}^{r}$ ($G^{r}$) denotes the retarded Green's function
in absence (presence) of random flux. Prior to disorder
averaging, two momentum labels are required for the Green's function $G^{r}$ in a disordered system, due to the breakdown of translation symmetry. Consequently,
the correction resulting from the diamagnetic effect is evaluated
as $\Sigma_{\mathrm{DM}}({\bf k})=\langle V({\bf k},{\bf k})\rangle$
yielding

\begin{equation}
\Sigma_{\mathrm{DM}}({\bf k})=\frac{\partial^{2}H_{0}({\bf k})}{\partial k_{\alpha}^{2}}\left(1-e^{-\frac{1}{2\mathcal{V}}\sum_{{\bf q}}D_{\alpha\alpha}({\bf q})}\right),
\end{equation}
where $H_{0}({\bf k})$ is the Hamiltonian in momentum space and $\mathcal{V}$
is the area of the system. We note that the exponential
form of this correction term indicates an infinite order of diamagnetic
terms. Incorporating these corrections, the model parameters in Eq.\ \eqref{eq:Model1}
are renormalized to

\begin{equation}
\tilde{b}_{\alpha}=b_{\alpha}e^{-\frac{1}{2\mathcal{V}}\sum_{{\bf q}}D_{\alpha\alpha}({\bf q})}
\end{equation}
and $\tilde{v}_{\alpha}=v_{\alpha}e^{-\frac{1}{2\mathcal{V}}\sum_{{\bf q}}D_{\alpha\alpha}({\bf q})}$.
This renormalization effectively renders electrons to be ``dressed''
by random flux. We observe that $\tilde{b}_{\alpha}$ decreases as
$U_{d}$ increases, which qualitatively accounts for renormalization
trends of effective masses at HSPs in Fig.\ \ref{fig:mass}(a).\textcolor{purple}{{}
}In the AWD model, the diamagnetic effect modifies the momentum-related
parameter $b_{\alpha}$ and velocity $v_{\alpha}$, ultimately altering
the band topology. In non-topological systems, the presence of diamagnetic term leads to an increasement in the total energy and consequently to a counteracting magnetic field.

Considering all diagrams with lowest-order non-crossing impurity lines,
it yields the self-consistent approximation of the self-energy {[}see
Fig.\ \ref{fig:self-energy}(b){]}:

\begin{align}
\Sigma(E,{\bf k}) & =\int_{BZ}\frac{d^{2}q}{(2\pi)^{2}}D_{\alpha\beta}({\bf q})\Gamma_{\alpha}({\bf k},{\bf k}-{\bf q})\times\nonumber \\
 & \frac{1}{E-\tilde{H}({\bf k}-{\bf q})-\Sigma(E,{\bf k}-{\bf q})}\Gamma_{\beta}({\bf k}-{\bf q},{\bf k}),
\end{align}
where $\tilde{H}({\bf k}-{\bf q})$ denotes the Hamiltonian modified
by the diamagnetic correction. The vertex term is defined as $\Gamma_{\alpha}({\bf k},{\bf k}')\equiv\frac{1}{2}[J_{\alpha}({\bf k})+J_{\alpha}({\bf k}')]$.
After incorporating the self-energy $\Sigma(E,{\bf k})$, it quantitatively
accounts for the modified phase boundaries of the AWD model in the
presence of random flux. We determine phase boundaries by imposing
gap closing condition at $\Gamma,X,Y$ and $M$. As shown in Fig.\ \ref{fig:phase1}(c),
the analytical phase boundary (dashed black lines) coincide with the
ones obtained numerically. Therefore, our
theory explains the topological phase transitions in terms of momentum-dependent
renormalization of model parameters, arising from the diamagnetic
effect and multiple scatterings off the random flux. We emphasize that this new
mechanism is clearly different from a uniform mass shift relevant for
the topological Anderson insulator \cite{Groth09prl}.

\section{Experimental implementation}
Our predictions can be tested in different physical platforms with high
tunability such as ultracold atoms, photonic crystals, and electric
circuits. Ultracold atoms are often able to nearly perfectly realize
topological models from theory \cite{Cooper19RMP}, where CIs have
been realized \cite{Jotzu14np,Liang23PRR}. The local magnetic field
for random flux can be achieved through lattice shading \cite{Hauke12prl,Goldman14RPP}
and laser-assisted tunneling \cite{Miyake13prl,Aidelsburger13prl,An17SA}.
The phase transition sequence can be directly detected by the evolution
of edge state configurations {[}Fig.\ \ref{fig:phase1}(d){]} and
corresponding transport signatures. Photonic crystals are also suitable
platforms for realizing different topological phases including CIs
\cite{Wang09Nature,LiC17prl,Ozawa19RRMP}, and the desired random
magnetic field can be mimicked by an optomechanical scheme \cite{Schmidt15Optica,Fang17NP,Aidelsburger18CRS}.
Moreover, our proposal can be realized in topological electric circuits
\cite{LeeCH18CP,DongJ21prr}, in which the random flux can be achieved
by tuning the complex hopping phases by designing the impedance network
\cite{Chen23NC}.

\section{Discussion and conclusion}
While chemical potential disorder has been extensively studied in topological systems, random flux induces fundamentally distinct phenomena that challenge conventional expectations. Our systematic comparison with prior results (see Table \ref{Table:Compare}) reveals that only the CI to NI transition follows established patterns. However, random flux drives unprecedented effects, notably revealing the role of weak topology in topological phase transitions and the emergence of new critical behavior.  

Our findings can be generalized to other topological systems. We expect similar physics in three-dimensional topological materials and systems with higher Chern numbers. Moreover, generalizing our approach from U(1) random flux to non-Abelian SU(2) gauge fields presents a promising direction for discovering novel disordered topological phases.

\begin{table}[h]
\centering
\caption{Comparison of our results from random flux with previous ones from chemical potential disorders. The previous results are referenced from Refs. \cite{Onoda03prl,Thonhauser08prb,Prodan10prl,Prodan11JPA,XueY13prb,Yushihito19prb,Moreno-Gonzalez23AP,Andrews24prb}.}
\label{table:Chernnumber-1}

\begin{tabular}{lcc}
\hline\hline
 & \makecell{Previous results} & \makecell{This research} \\
\midrule 
CI$\rightarrow$NI  & \textbf{\LARGE$\checkmark$} & \textbf{\LARGE$\checkmark$} \\
\midrule 
NI$\rightarrow$CI & \textbf{\LARGE$\checkmark$} & \textbf{\LARGE$\times$} \\
\midrule 
Weak topology & \textbf{\LARGE$\times$} & \textbf{\LARGE$\checkmark$} \\
\midrule 
WTI-x$\rightarrow$CI  & \textbf{\LARGE$\times$} & \textbf{\LARGE$\checkmark$} \\
\midrule 
CI$\rightarrow$WTI-y  & \textbf{\LARGE$\times$} & \textbf{\LARGE$\checkmark$} \\
\midrule 
Emergent phase & \textbf{\LARGE$\times$} & \textbf{\LARGE$\checkmark$} \\
\midrule 
\makecell{New quantum criticality} & \makecell{\textbf{\LARGE$\times$}} & \makecell{\textbf{\LARGE$\checkmark$}} \\
\hline\hline
\end{tabular}\label{Table:Compare}
\end{table}

In summary, we study the fundamental
and important role that random flux plays in topological phases of
matter. Based on the AWD model in 2D, we have demonstrated a random-flux-induced
phase transition sequence WTI-x$\rightarrow$CIs$\rightarrow$WTI-y
with the reentrance of a different type of WTI. The anisotropy and
weak topology of the system result in the emergence of quasi-critical
points at the phase transitions, characterized by a spatially anisotropic
localization behavior and a new critical exponent $\nu\simeq5.46$.
We provide consistent analysis of the phase transitions and
quantum critical phenomena based on numerics and analytical modelling. Our results reveal a qualitatively
new disorder effect due to random flux in topological systems.

\section{Acknowledgement}
We thank Shun-Qing Shen and Jan Budich for helpful discussion. C.A.L. thanks N. Bauer and G. Starkov for help on supercomputer
cluster resources. This work was supported by the DFG (SFB 1170 Project-Id:
258499086), and the Würzburg-Dresden Cluster of Excellence ct.qmat,
EXC 2147 (Project-Id: 390858490). We thank the Bavarian Ministry of
Economic Affairs, Regional Development and Energy for financial support
within the High-Tech Agenda Project \textquotedblleft Bausteine für
das Quanten Computing auf Basis topologischer Materialen.\textquotedblright{}
B.F. is financially supported by Guangdong Basic and Applied Basic
Research Foundation No. 2024A1515010430 and No. 2023A1515140008. J.L.
acknowledges the support from NSFC under Project No. 92265201 and
the Innovation Program for Quantum Science and Technology under Project
No. 2021ZD0302704.

\appendix

\section{Energy spectrum of the anisotropic Wilson-Dirac model}
In this appendix, we present the band structure, Chern number, and
weak topology of the anisotropic Wilson-Dirac (AWD) model.

\begin{figure*}[t]
\includegraphics[width=1\linewidth]{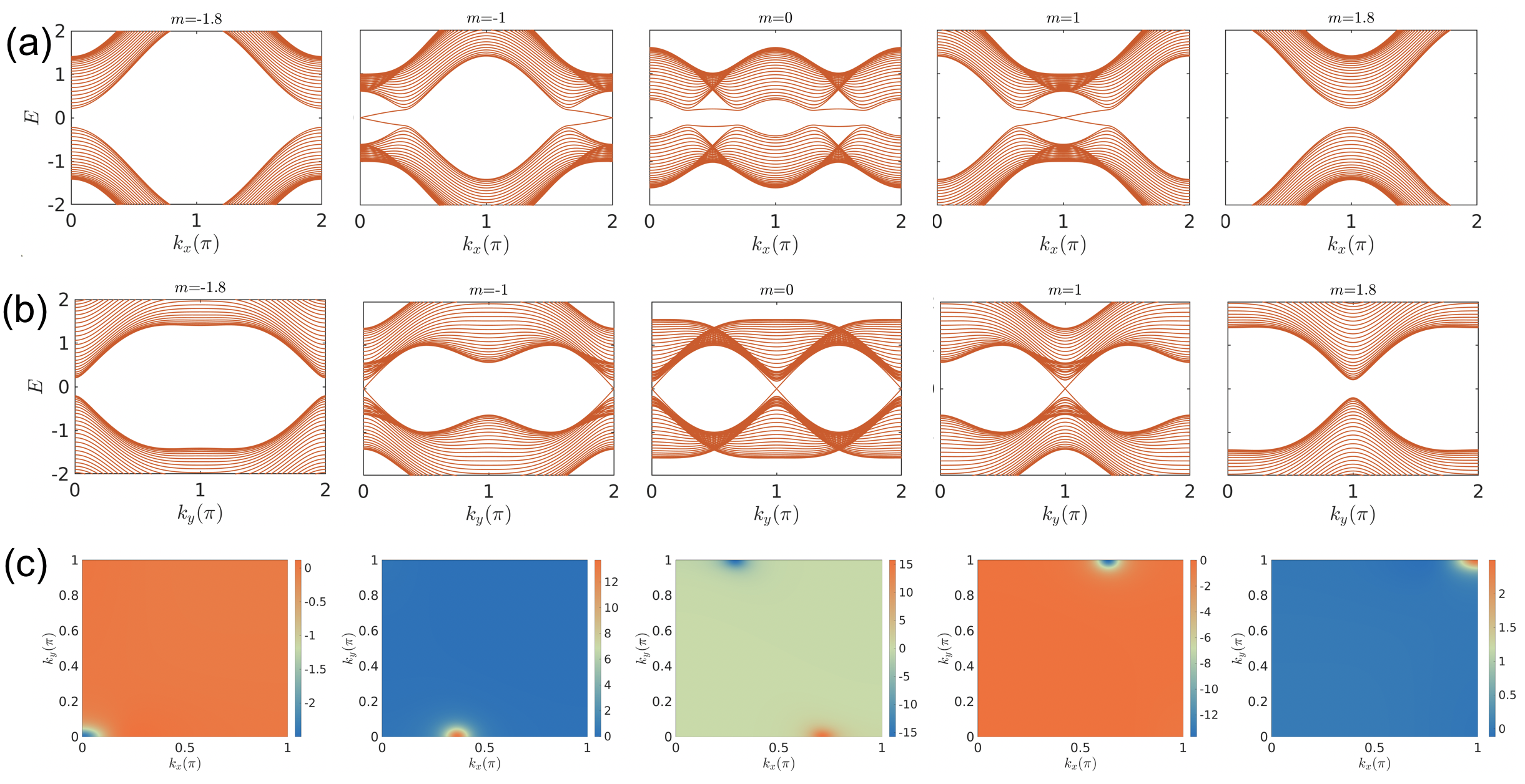}

\caption{(a) Energy spectra of a ribbon geometry along $x$ direction under
open boundary conditions for different parameter $m$. (b) The same
as (a) but along $y$ direction. (c) The Berry curvature $F(k_{x},k_{y})$
plotted corresponding the parameter settings in (a) and (b). Other
parameters are: $b_{x}=1,b_{y}=0.6,v_{x}=0.2$, and $v_{y}=1$. \label{fig:bands}}
\end{figure*}

The nonzero Chern number indicates the existence of chiral edge modes
at open boundaries. In Figs.\ \ref{fig:bands}(a) and \ref{fig:bands}(b),
we show the energy spectra along $x$ and $y$ directions, respectively,
corresponding to five different phase regions with increasing the
parameter $m$ in Fig. 2(a) of the main text. Let us focus on the
case with $m=0$. Here, the Chern number is $C=0$ since the Berry
curvature cancels out across the Brillouin zone {[}Fig.\ \ref{fig:bands}(c){]}.
However, we observe two pairs of Dirac edge modes along $y$ direction
(at $k_{y}^{*}=0,\pi$) but no Dirac edge modes along $x$ direction.
This phase is identified as a weak topological insulator (WTI). Therefore,
the Chern number $C=0$ cannot distinguish between the normal insulator
(NI) and WTI.

\section{Analytical theory for random-flux-induced topological phase transitions}
In this appendix, we present the details for the analytical
theory that accounts for random-flux-induced topological phase transitions
arising from the diamagnetic effect and multiple scattering of random
flux. The random flux is connected to a random magnetic field as
$\phi({\bf r})=B({\bf r})$. Therefore, the random magnetic field
is then uniformly distributed in the interval $[-U_{d}/2,U_{d}/2]$,
where $U_{d}$ represents the random flux strength. The disorder averages
of products of two random magnetic fields are given by

\begin{equation}
\langle B({\bf q})B(-{\bf q})\rangle=\frac{U_{d}^{2}}{12}\delta_{{\bf q}+{\bf q}',0},
\end{equation}
where $B({\bf q})$ is the Fourier transform of the random magnetic
field defined as $B({\bf q})=\int d{\bf r}e^{-i{\bf q}\cdot{\bf r}}B({\bf r})$.
The corresponding vector potential $A_{\alpha}({\bf q})$, which generates
the fluctuating magnetic field, can be expressed as

\begin{equation}
A_{\alpha}({\bf q})=\frac{i\epsilon_{\alpha\beta}q_{\beta}}{q^{2}}B({\bf q}),
\end{equation}
where $\epsilon_{\alpha\beta}$ is the Levi-Civita symbol in two dimensions,
and $\alpha,\beta$ range over $x,y$. The correlations between vector
potentials are given by 

\begin{equation}
\langle A_{\alpha}({\bf q})A_{\beta}({\bf q}')\rangle=\mathcal{V}\frac{U_{d}^{2}}{12}\frac{\delta_{\alpha\beta}-\hat{q}_{\alpha}\hat{q}_{\beta}}{|{\bf q}|^{2}}\delta_{{\bf q}+{\bf q}',0},\label{eq:ACorrelator}
\end{equation}
where $\mathcal{V}$ is the volume. For convenience, we define the
correlator 

\begin{equation}
D_{\alpha\beta}({\bf q})\equiv\frac{U_{d}^{2}}{12}\frac{\delta_{\alpha\beta}-\hat{q}_{\alpha}\hat{q}_{\beta}}{|{\bf q}|^{2}},
\end{equation}
where the term $(\delta_{\alpha\beta}-\hat{q}_{\alpha}\hat{q}_{\beta})$
accounts for the transverse nature of the vector potential with $\hat{q}_{\alpha}\equiv q_{\alpha}/|{\bf q}|$. 

In the presence of random flux, the minimally coupled Hamiltonian
$\hat{H}_{0}$ is transformed to $\hat{H}_{A}$ given by
\begin{equation}
\hat{H}_{A}=\sum_{{\bf r}}\hat{\psi}_{{\bf r}}^{\dagger}{\bf d}[-i\partial_{{\bf r}}+e{\bf A}({\bf r})]\cdot\bm{\sigma}\hat{\psi}_{{\bf r}}.
\end{equation}
We expand the full Hamiltonian into two components 

\begin{equation}
\hat{H}_{A}=\hat{H}_{0}+\hat{V}[{\bf A}({\bf r})],
\end{equation}
where $\hat{H}_{0}$ is the Hamiltonian in the clean limit and $\hat{V}[{\bf A}({\bf r})]$
is the external potential that can be expressed as $\hat{V}=\sum_{{\bf k},{\bf k}'}\psi_{{\bf k}}^{\dagger}V({\bf k},{\bf k}')\psi_{{\bf k}'}$.
Here, $V({\bf k},{\bf k}')$ includes interaction terms of all orders
\begin{equation}
V({\bf k},{\bf k}')=\sum_{n=1}^{\infty}V_{n}({\bf k},{\bf k}'),\label{eq:PerturbPotential}
\end{equation}
where $V_{n}({\bf k},{\bf k}')$ accounts for the $n$-th order interaction
with explicit form 

\begin{align}
V_{n}({\bf k},{\bf k}') = \frac{1}{n!}\sum_{{\bf q}_{1},{\bf q}_{2},\cdots,{\bf q}_{n-1}}\Gamma_{\bm{\alpha}}({\bf k},{\bf k}')A_{\alpha_{1}}(-{\bf q}_{1}) \nonumber \\ A_{\alpha_{2}}(-{\bf q}_{2})\cdots 
A_{\alpha_{n}}({\bf k}-{\bf k}'+{\bf q}_{n}),
\end{align}
where ${\bf q}_{n}\equiv\sum_{i=1}^{n-1}{\bf q}_{i}$, $\bm{\alpha}\equiv\{\alpha_{1},\alpha_{2},\cdots,\alpha_{n}\}$
and $\Gamma_{{\bf \bm{\alpha}}}({\bf k},{\bf k}')=\frac{1}{2}[J_{\bm{\alpha}}({\bf k})+J_{\bm{\alpha}}({\bf k'})]$
is the $n$-th order vertex. $J_{\bm{\alpha}}({\bf k})$ is the trapezoidal
current operator defined as

\begin{equation}
J_{\bm{\alpha}}({\bf k})\equiv\frac{\partial_{n}H({\bf k})}{\partial k_{\alpha_{1}}\partial k_{\alpha_{2}}...\partial k_{\alpha_{n}}}.
\end{equation}

\begin{widetext}
The random flux averaging of the external term $\hat{V}[A({\bf r})]$
involves multiple orders of the vector potential, which are taken
into account by introducing the generating functional

\begin{align}
\mathcal{G}[\eta,\bar{\eta}] & =\mathcal{Z}^{-1}\int D[A]e^{-\frac{1}{2}\int d{\bf r}\int d{\bf r}'A_{\alpha}({\bf r})D_{\alpha\beta}^{-1}({\bf r},{\bf r}')A_{\beta}({\bf r}')+\int d{\bf r}\eta_{\alpha}({\bf r})A_{\alpha}({\bf r})}\nonumber \\
 & =e^{\int d{\bf r}\int d{\bf r}'\eta_{\alpha}({\bf r})D_{\alpha\beta}({\bf r}-{\bf r}')\eta_{\beta}({\bf r}')},
\end{align}
where $\eta_{\alpha}({\bf r})$ is an arbitrary real field, and $\mathcal{Z}=\int D[A]e^{-\frac{1}{2}\int d{\bf r}\int d{\bf r}'A_{\alpha}({\bf r})D_{\alpha\beta}^{-1}({\bf r},{\bf r}')A_{\beta}({\bf r}')}$
is the multidimensional Gaussian integral. Differentiating the functional
integral twice with respect to $\eta$ according to $\frac{\partial^{2}\mathcal{G}[\eta,\bar{\eta}]}{\partial\eta_{\alpha}({\bf r})\partial\eta_{\beta}({\bf r}')}|_{\eta,\bar{\eta}=0}$
yields the averaged products of two vector potentials 

\begin{equation}
\langle A_{\alpha}({\bf r})A_{\beta}({\bf {\bf r}}')\rangle=D_{\alpha\beta}({\bf r}-{\bf r}'),
\end{equation}
which is exactly the Fourier transformation of the Eq.\ \eqref{eq:ACorrelator}:
$D_{\alpha\beta}({\bf r}-{\bf r}')=\frac{1}{\mathcal{V}}\sum_{{\bf q}}e^{i{\bf q}\cdot({\bf r}-{\bf r}')}D_{\alpha\beta}({\bf q})$.
For higher-order averages involving $2n$ vector potentials, differentiation
of the generating function $2n$ times results in 

\begin{align}
\langle A_{\alpha_{1}}({\bf r}_{1})A_{\alpha_{2}}({\bf r}_{2})\cdots A_{\alpha_{2n}}({\bf r}_{2n})\rangle= \sum_{\mathrm{pairs\ of}\{i_{1},\cdots,i_{2n}\}}
D_{\alpha_{i_{1}}\alpha_{i_{2}}}({\bf r}_{\alpha_{i_{1}}}-{\bf r}_{\alpha_{i_{1}}})\times\cdots D_{\alpha_{i_{2n-1}}\alpha_{i_{2n}}}({\bf r}_{\alpha_{i_{2n-1}}}-{\bf r}_{\alpha_{i_{2n}}}),\label{eq:APairs}
\end{align}
which is given by all possible pairings that can be formed from the
$2n$ components of $A({\bf r})$.

We now evaluate the random flux averaging of the external term as
defined in Eq.\ \eqref{eq:PerturbPotential} 
\begin{equation}
\langle V({\bf k},{\bf k}')\rangle=\sum_{n=1}^{\infty}\frac{1}{n!\mathcal{V}^{n}}\sum_{{\bf q}_{1},{\bf q}_{2},\cdots,{\bf q}_{n-1}}\Gamma_{\bm{\alpha}}({\bf k},{\bf k}')\langle A_{\alpha_{1}}(-{\bf q}_{1})A_{\alpha_{2}}(-{\bf q}_{2})\cdots A_{\alpha_{n}}({\bf k}-{\bf k}'+{\bf q}_{n})\rangle.
\end{equation}
According to Eq.\ \eqref{eq:APairs}, only even orders of $A$ survive
after disorder averaging (see Fig. 4(a) in the main text). Given
that $V({\bf k},{\bf k}')$ contains all order of vector potentials,
the random configuration average at the first order $\langle V({\bf k},{\bf k}')\rangle$
is nonzero, in sharp contrast to that of on-site potential disorder.
Using the relations $\frac{1}{\mathcal{V}}\sum_{{\bf q}}D_{\alpha\beta}({\bf q})=\delta_{\alpha\beta}\frac{1}{\mathcal{V}}\sum_{{\bf q}}D_{\alpha\alpha}({\bf q})$
and $\Gamma_{[\alpha\cdots\alpha]_{2n}}({\bf k},{\bf k})=(-1)^{n+1}J_{\alpha\alpha}({\bf k})$
for the AWD model, it can be further recast in a closed form as

\begin{equation}
\langle V({\bf k},{\bf k}')\rangle=J_{\alpha\alpha}({\bf k})[1-e^{-\frac{1}{2\mathcal{V}}\sum_{{\bf q}}D_{\alpha\alpha}({\bf q})}]\delta_{{\bf k},{\bf k}'}.\label{eq:Vaverage}
\end{equation}

We then evaluate the random flux averaging over the product of two
external terms, which involves considering all combinations of two
$V({\bf k},{\bf k}')$ as 
\begin{align}
\langle V({\bf k},{\bf k}_{1})\otimes V({\bf k}_{1},{\bf k}')\rangle\nonumber \\
= & \frac{1}{\mathcal{V}^{2}}\Gamma_{\alpha}({\bf k},{\bf k}_{1})\otimes\Gamma_{\alpha'}({\bf k}_{1},{\bf k}')\langle A_{\alpha}({\bf k}-{\bf k}_{1})A_{\alpha'}({\bf k}_{1}-{\bf k}')\rangle\nonumber \\
+ & \frac{1}{3!\mathcal{V}^{4}}\Gamma_{\alpha}({\bf k},{\bf k}_{1})\otimes\Gamma_{\alpha'_{1}\alpha'_{2}\alpha'_{3}}({\bf k}_{1},{\bf k}')\sum_{{\bf q}'_{1},{\bf q}_{2}'}\langle A_{\alpha}({\bf k}-{\bf k}_{1})A_{\alpha_{1}'}(-{\bf q}_{1}')A_{\alpha_{2}'}(-{\bf q}_{2}')A_{\alpha_{3}'}({\bf k}_{1}-{\bf k}'+{\bf q}_{1}'+{\bf q}_{2}')\rangle\nonumber \\
+ & \frac{1}{3!\mathcal{V}^{4}}\Gamma_{\alpha{}_{1}\alpha{}_{2}\alpha{}_{3}}({\bf k},{\bf k}_{1})\otimes\Gamma_{\alpha'}({\bf k}_{1},{\bf k}')\sum_{{\bf q}{}_{1},{\bf q}_{2}}\langle A_{\alpha_{1}}(-{\bf q}_{1})A_{\alpha_{2}}(-{\bf q}_{2})A_{\alpha_{3}}({\bf k}_{1}-{\bf k}+{\bf q}_{1}+{\bf q}_{2})A_{\alpha'}({\bf k}_{1}-{\bf k}')\rangle\nonumber \\
+ & \frac{1}{2!\mathcal{V}^{2}}\Gamma_{\alpha{}_{1}\alpha{}_{2}}({\bf k},{\bf k}_{1})\otimes\Gamma_{\alpha_{1}'\alpha{}_{2}'}({\bf k}_{1},{\bf k}')\sum_{{\bf q}'_{1},{\bf q}_{2}'}\langle A_{\alpha_{1}}(-{\bf q}_{1})A_{\alpha_{2}}({\bf k}-{\bf k}_{1}+{\bf q}_{1})A_{\alpha_{1}'}(-{\bf q}_{1}')A_{\alpha'}({\bf k}_{1}-{\bf k}'+{\bf q}_{1}')\rangle\nonumber \\
+ & \cdots
\end{align}
By truncating to the lowest order of nonvanishing terms, we approximate:
\begin{equation}
\langle V({\bf k},{\bf k}_{1})\otimes V({\bf k}_{1},{\bf k}')\rangle\simeq\frac{1}{\mathcal{V}}\Gamma_{\alpha}({\bf k},{\bf k}_{1})\otimes\Gamma_{\alpha'}({\bf k}_{1},{\bf k}')D_{\alpha\alpha'}({\bf k}-{\bf k}')\delta_{{\bf k}{\bf k}'}.\label{eq:Vaverage2}
\end{equation}

The Dyson equation in the presence of the $V({\bf k},{\bf k}')$ can
be expressed as 
\begin{align}
G^{r}({\bf k},{\bf k}',E) & =\delta_{{\bf k}{\bf k}'}G_{0}^{r}({\bf k},E)+G_{0}^{r}({\bf k},E)\sum_{{\bf k}''}V({\bf k},{\bf k}'')G^{r}({\bf k}'',{\bf k}',E).
\end{align}
Prior to impurity averaging, the Green\textquoteright s function requires
two momentum labels because translation symmetry is broken. To handle
this, one can iteratively solve the right-hand side of the Dyson equation,
averaging terms sequentially using Eqs.\ \eqref{eq:Vaverage} and
\eqref{eq:Vaverage2}. This process helps to identify repeating structures
that can be summed to an infinite order: 
\begin{alignat}{1}
\langle G^{r}({\bf k},{\bf k}',E)\rangle & =\delta_{{\bf k}{\bf k}'}G_{0}^{r}({\bf k},E)+G_{0}^{r}({\bf k},E)\langle V({\bf k},{\bf k}')\rangle G_{0}^{r}({\bf k}',E)\nonumber \\
 & +G_{0}^{r}({\bf k},E)\sum_{{\bf k}_{1}}\langle V({\bf k},{\bf k}_{1})G_{0}^{r}({\bf k}_{1},E)V({\bf k}_{1},{\bf k}')\rangle G_{0}^{r}({\bf k}',E)+\cdots\nonumber \\
 & =\delta_{{\bf k}{\bf k}'}G_{0}^{r}({\bf k},E)\{1+\langle V({\bf k},{\bf k}')\rangle G_{0}^{r}({\bf k}',E)+\langle V({\bf k},{\bf k})\rangle G_{0}^{r}({\bf k},E)\langle V({\bf k},{\bf k})\rangle G_{0}^{r}({\bf k},E)\nonumber \\
 & +\frac{1}{\mathcal{V}}\sum_{{\bf k}_{1}}D_{\alpha\alpha'}({\bf k}-{\bf k}_{1})\Gamma_{\alpha}({\bf k},{\bf k}_{1})G_{0}^{r}({\bf k}_{1},E)\Gamma_{\alpha'}({\bf k}_{1},{\bf k})G_{0}^{r}({\bf k},E)+\cdots\}.
\end{alignat}
After disorder averaging, translational symmetry is restored, as indicated
by the presence of $\delta_{{\bf k}{\bf k}'}$. We can sum infinite
subsets of diagrams, as depicted in Fig. 4 of the main text. The full
Green\textquoteright s function is then expressed as:

\begin{equation}
G^{r}({\bf k},E)=[(G_{0}^{r}({\bf k},E))^{-1}-\Sigma({\bf k},E)]
\end{equation}
with the self-energy given by
\begin{equation}
\Sigma({\bf k},E)=\langle V({\bf k},{\bf k})\rangle+\frac{1}{\mathcal{V}}\sum_{{\bf k}_{1}}D_{\alpha\alpha'}({\bf k}-{\bf k}_{1})\Gamma_{\alpha}({\bf k},{\bf k}_{1})G^{r}({\bf k}_{1},E)\Gamma_{\alpha'}({\bf k}_{1},{\bf k}).
\end{equation}
The self-energy shows explicit dependence on wave vector ${\bf k}$,
distinguishing it clearly from the on-site potential disorder case
with a uniform mass renormalization \citep{Groth09prl}. After incorporating
the self-energy $\Sigma(E,{\bf k})$, the preceding approach can quantitatively
account for the modified phase diagram of the AWD model in the presence
of random flux, as discussed in the main text.

\begin{figure}
\includegraphics[width=0.5\linewidth]{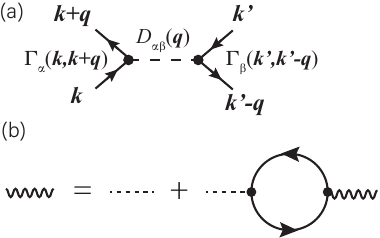}

\caption{Effective four-fermion scatterings induced by the random flux. (b)
Renormalization of the random gauge field correlator due to electron-field
interactions. Wavy lines represent the dressed correlator for the
vector potential. Dashed lines depict the bare correlator for the
vector potential. Solid lines with arrows denote the electron propagators.
\label{fig:FeynmanD}}
\end{figure}

Note that in the above calculations, the factor $D_{\alpha\beta}({\bf k}-{\bf k}')$
exhibits a strong singularity in the forward direction ${\bf k}-{\bf k}'$
due to the extensive range of the vector potential fluctuations, despite
the magnetic field fluctuations are short-ranged. To manage the infrared
divergences in the self-energy calculations, we utilize the regulated
correlator $\tilde{D}_{\alpha\beta}({\bf k}-{\bf k}')$ instead. Drawing
an analogy from the correction of the polarization tensor to the free
gauge boson propagator in quantum electrodynamics (QED), we derive
higher-order corrections to the impurity correlator for a given system.
This involves evaluating a diagram analogous to those used in QED
as shown in Fig.\ \ref{fig:FeynmanD} and solving a self-consistent
equation for the renormalized impurity correlator:

\begin{equation}
\tilde{D}_{\alpha\beta}({\bf q})=D_{\alpha\beta}({\bf q})+D_{\alpha\gamma}({\bf q})\Pi_{\gamma\delta}({\bf q})\tilde{D}_{\alpha\beta}({\bf q}),\label{eq:Impuritycorre}
\end{equation}
where the polarization is given by 

\begin{equation}
\Pi_{\gamma\delta}({\bf q})=\frac{1}{\mathcal{V}}\sum_{{\bf k}}\mathrm{Tr}[\Gamma_{\gamma}({\bf k}+{\bf q},{\bf k})G^{r}({\bf k})\Gamma_{\delta}({\bf k},{\bf k+q})G^{r}({\bf k}+{\bf q})].
\end{equation}
It is revealed that the absolute values off-diagonal elements $\{|\Pi_{xy}|,|\Pi_{yx}|\}$
are much smaller than the diagonal elements $\{|\Pi_{xx}|,|\Pi_{xy}|\}$
with $\Pi_{xx},\Pi_{xy}<0$. Then, the polarization tensor can be
approximated as $\Pi=\mathrm{diag}\{\Pi_{xx},\Pi_{xy}\}$. By substituting
this result into Eq.\ \eqref{eq:Impuritycorre}, we find 

\begin{equation}
\tilde{D}_{\alpha\beta}({\bf q})\equiv\frac{U_{d}^{2}}{12}\frac{\delta_{\alpha\beta}-\hat{q}_{\alpha}\hat{q}_{\beta}}{|{\bf q}|^{2}+\lambda_{s}^{-2}},
\end{equation}
where the screening length is defined as $\lambda_{s}=1/\sqrt{-\frac{U_{d}^{2}}{12}(\Pi_{yy}\hat{q}_{x}^{2}+\Pi_{xx}\hat{q}_{y}^{2})}$.
Consequently, the correlator of the random vector potential acquires
effective screening due to the electron-gauge field interaction.

\end{widetext}

\section{Phase diagram under different parameter settings}
\begin{figure}
\includegraphics[width=1.0\linewidth]{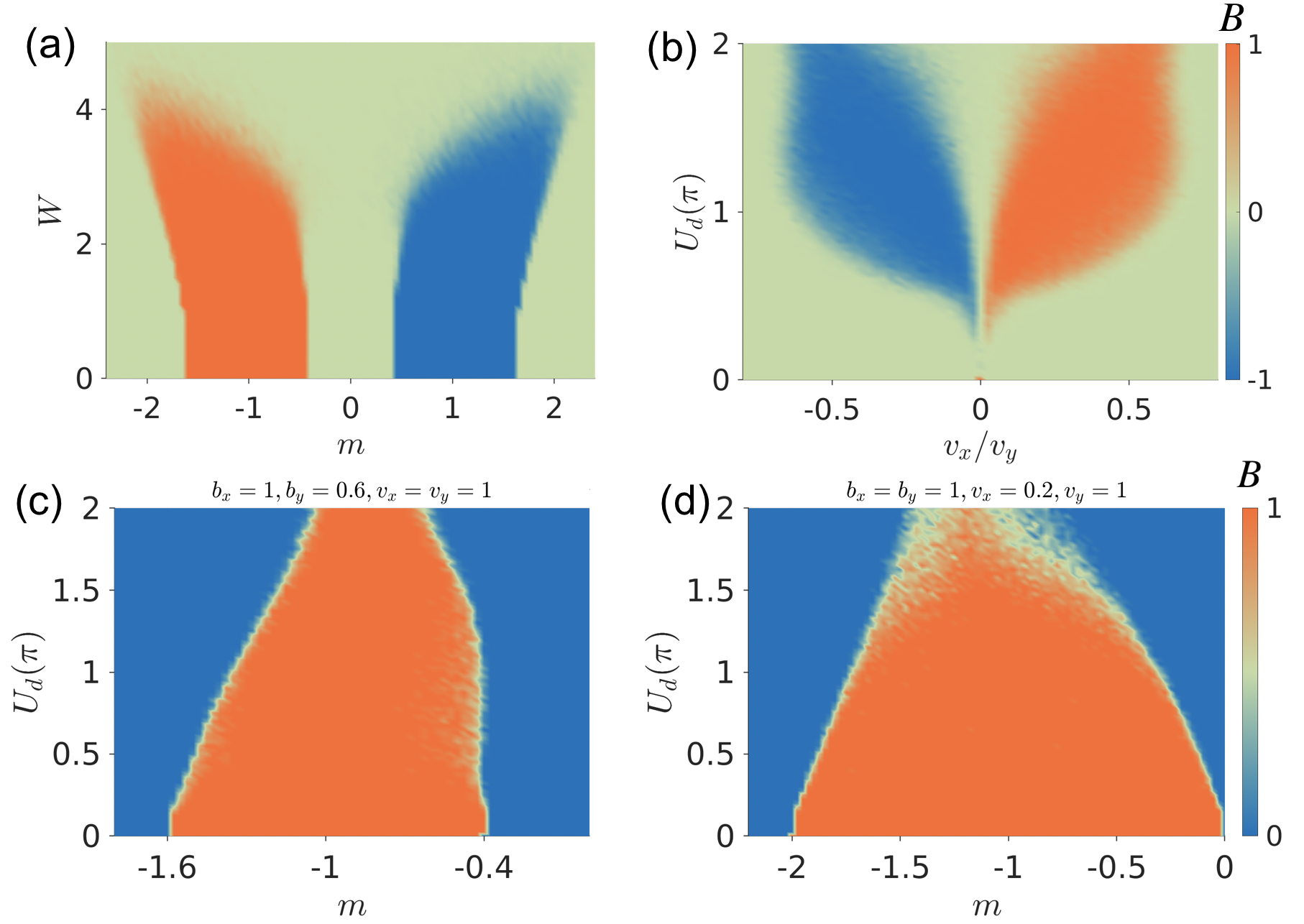}

\caption{Phase diagram of the AWD model under different parameter settings.
(a) Bott index as functions of on-site disorder strength $W$ and
mass $m$. No random flux is applied. Other parameters are: $b_{x}=1,b_{y}=0.6,v_{x}=0.2$,
and $v_{y}=1$. (b) Bott index as as a function random flux strength
$U_{d}$ and velocity ratio $v_{x}/v_{y}$ for fixed mass term $m=-0.3$.
Other parameters are $b_{x}=1$ and $b_{y}=0.6$. (c) Bott index as
a function $U_{d}$ and $m$ for the condition of $b_{x}\protect\neq b_{y}$
and $v_{x}=v_{y}$. (d) The same as (c) but with the condition of
$b_{x}=b_{y}$ and $v_{x}\protect\neq v_{y}$. \label{fig:phase_iso}}
\end{figure}

In this appendix, we present the phase diagrams in terms of the Bott
index under different parameter settings. Figure \ref{fig:phase_iso}(a)
illustrates the phase diagram for the case of on-site potential disorder,
serving as a comparison with the random flux scenario. The on-site
disorder takes a form of $V({\bf r})I_{2\times2}$, where $V({\bf r})$
is uniformly distributed in the interval $[-W/2,W/2]$, with $W$
denoting the disorder strength. We find that the phase boundaries
incline in opposite directions compared to the random flux case as
shown in the Fig. 2(c) of the main text. Here, the topological Anderson
transition from NIs to CIs happens as expected. 

We note that the anisotropy from Fermi velocity ratio $v_{x}/v_{y}$
also plays an important rule in these transitions. As illustrated
in Fig.\ \ref{fig:phase_iso}(b), the system undergoes topological
phase transitions only within a proper regime of $v_{x}/v_{y}$. Such
a sensitive dependence on Fermi velocity ratio is absent in the on-site
disorder scenario. If we set $v_{x}=v_{y}$ but $b_{x}\neq b_{y}$
in Fig.\ \ref{fig:phase_iso}(c), the random flux-induced CIs will
not appear anymore. While for the case $b_{x}=b_{y}$, there is no
WTI phase in the first place, and the CI is driven to a NI as increasing
$U_{d}$ {[}Fig.\ \ref{fig:phase_iso}(d){]}.

\section{Effective band structures from averaged Green's function}
In this appendix, we present the effective band structures obtained
from averaged Green's function and their corresponding Berry curvatures.
We average the Green's function over a large enough number of random
configurations, such that the translation symmetry can be effectively
restored and an effective Hamiltonian can be obtained. 

\begin{figure}
\includegraphics[width=1\linewidth]{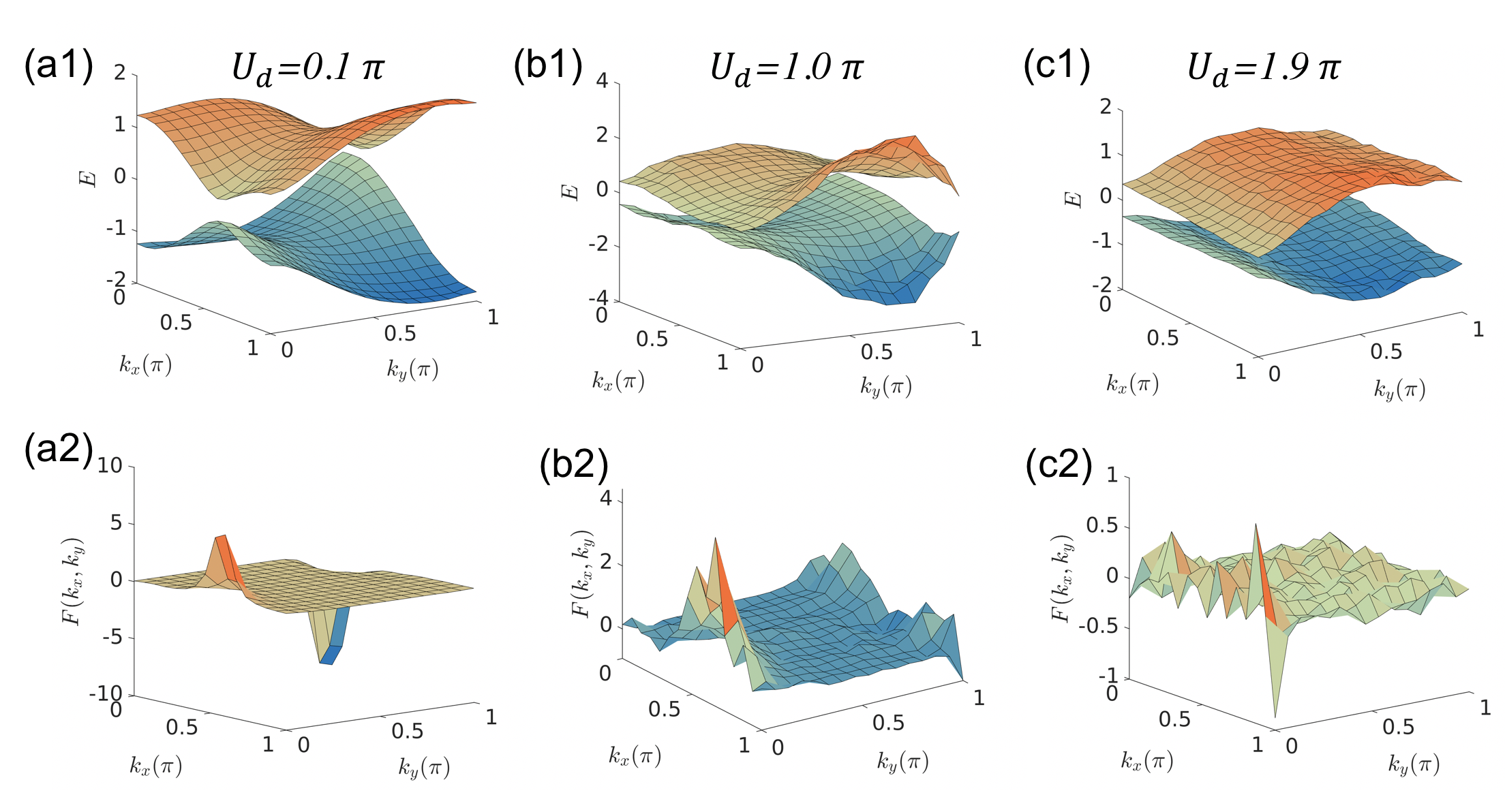}

\caption{Effective bands structures {[}upper panel{]} and their corresponding
Berry curvatures {[}lower panel{]} for representative random flux
strength $U_{d}=0.1\pi,1.0\pi$, and $1.9\pi$, respectively. Other
parameters are: $m=-0.3,b_{x}=1,b_{y}=0.6,v_{x}=0.2$, and $v_{y}=1$.
\label{fig:effBands}}
\end{figure}

The random flux averaged Green's function is given by $G_{\mathrm{avg}}^{r}({\bf r}-{\bf r}',E)=\langle G^{r}({\bf r},{\bf r}',E)\rangle$,
where $\langle...\rangle$ indicates the disorder average. By Fourier
transforming the averaged Green's function, we obtain $G^{r}({\bf k},E)=\sum_{{\bf r}}G_{\mathrm{avg}}^{r}({\bf r},E)e^{i{\bf k}\cdot{\bf r}}$.
From this, the effective Hamiltonian is derived as 

\begin{equation}
H_{\mathrm{eff},G}({\bf k})=-[G^{r}({\bf k},E=0)]^{-1}.
\end{equation}

The Berry curvatures can be calculated by the lattice gauge theory
method described in Ref. \citep{Yoshimura14prb}. First, we solve
the eigenvalue problem for the effective Hamiltonian $H_{\mathrm{eff},G}({\bf k})$
on a discretized Brillouin zone mesh, $H_{\mathrm{eff},G}({\bf k}_{j})|\varphi_{n}({\bf k}_{j})\rangle=\epsilon_{n}|\varphi_{n}({\bf k}_{j})\rangle$,
where the momentum points are ${\bf k}_{j}\equiv(j_{x}e_{k_{x}},j_{y}e_{k_{y}})$
with $e_{k_{x}}=\frac{2\pi}{N_{x}}$ and $e_{k_{y}}=\frac{2\pi}{N_{y}}.$
The $U(1)$ link for the occupied band is defined as $M_{\alpha=x,y}({\bf k}_{j})\equiv|\mathrm{det}U_{\alpha}({\bf k}_{j})|^{-1}\mathrm{det}U_{\alpha}({\bf k}_{j})$
with the matrix $U_{\alpha}({\bf k}_{j})=\langle\varphi({\bf k}_{j})|\varphi({\bf k}_{j}+\hat{e}_{\alpha})\rangle$.
This link variables are well-defined except at singular points with
$\mathrm{det}\ U_{\alpha}({\bf k}_{j})=0$. Using these link variables,
we obtain a lattice field strength as
\begin{equation}
F({\bf k}_{j})\equiv\mathrm{ln}\left[M_{x}({\bf k}_{j})M_{y}({\bf k}_{j}+\hat{e}_{x})M_{x}^{-1}({\bf k}_{j}+\hat{e}_{y})M_{y}^{-1}({\bf k}_{j})\right].
\end{equation}

In Figs.\ \ref{fig:effBands}(a1), \ref{fig:effBands}(b1), and\ \ref{fig:effBands}(c1),
we show the band structures of the effective Hamiltonian $H_{\mathrm{eff},G}({\bf k})$
for different random flux strengths, $U_{d}=0.1\pi,1.0\pi$, and $1.9\pi$,
respectively. The band structure evolves as changing $U_{d}$. Figures
\ref{fig:effBands}(a2), \ref{fig:effBands}(b2), and \ref{fig:effBands}(c2)
are the corresponding Berry curvatures $F(k_{x},k_{y})$. For small
random flux strength $U_{d}=0.1\pi$, the Berry curvature features
both peak and dip, which cancel out, resulting in a Chern number $C=0$.
It is still in the WTI-x with $(C;w_{x}w_{y})=(0;10)$. At an intermediate
random flux strength $U_{d}=1.0\pi$, the system is driven to a Chern
insulator with $C=1$. The corresponding Berry curvature show peaks
only {[}Fig. \ref{fig:effBands}(b2){]}. As increasing $U_{d}$ further
to $U_{d}=1.9\pi$, the Berry curvature gives zero Chern number and
the phase WTI-y phase with $(C;w_{x}w_{y})=(0;01)$.

\section{Critical features at the phase transition points from CIs to NIs}
\begin{figure}
\includegraphics[width=1.0\linewidth]{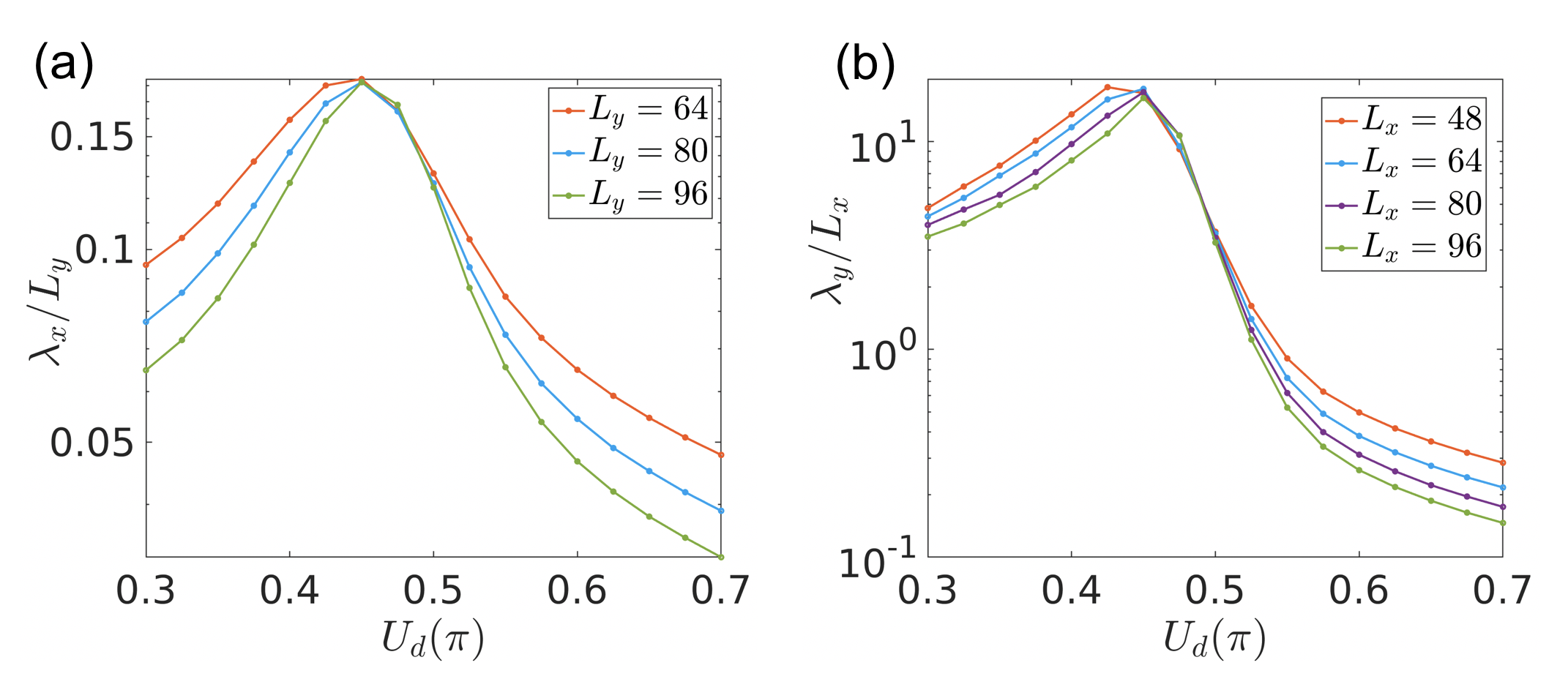}

\caption{Localization length on a quasi-1D tube as a function of $U_{d}$.
(a): along $x$ direction; (b): along $y$ direction. Other parameters
are: $b_{x}=1,b_{y}=0.6,v_{x}=0.2$, $v_{y}=1$, and $m=-1.5$. \label{fig:scaling_critical}}
\end{figure}

In this section, we present the critical scaling signatures at the
phase transition points, specifically focusing on the transitions
from a CI to a NI as a comparison. As already shown in Fig.\ \ref{fig:transport},
the phase transition from a CI to a NI leads to the jump of Chern
number from $C=1$ to $C=0$, accompanied by a transition in conductance
from $G=\frac{e^{2}}{h}$ to $G=0$. During the topological phase
transition, a critical point emerges where the localization length
diverges. To investigate this, we present the scaling behavior of
normalized localization lengths as a function of random flux strength
$U_{d}$ in Fig.\ref{fig:scaling_critical}. We find that the
critical points appear both along $x$ direction [Fig.\ref{fig:scaling_critical}(a)]
and $y$ direction [Fig.\ref{fig:scaling_critical}(b)] at
around $U_{d}\simeq0.48\pi$, indicating the divergence of localization
lengths along both $x$ and $y$ directions. This is attributed to
the NI being trivial along both directions while CIs are nontrivial
along both directions. These results show stark contrast with the
quasi-critical points between the transitions of CIs and WTIs, where
the localization length diverges only along one of the two directions. 

\section{Random-flux-induced phase transitions in Haldane model}

\begin{figure}
\includegraphics[width=1.0\linewidth]{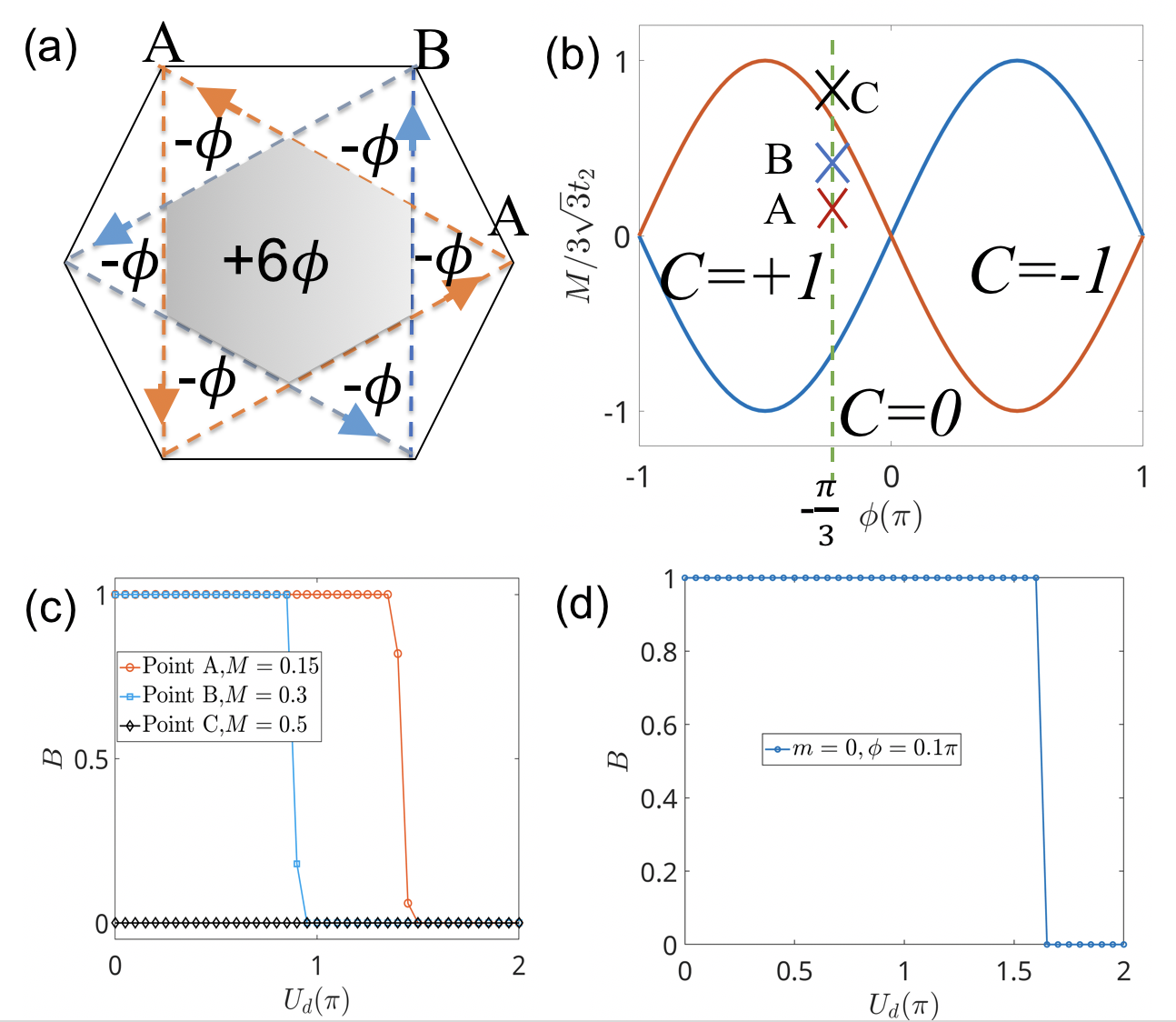}

\caption{(a) Haldane model on honeycomb lattice. Without random
flux, the next nearest-neighbor hopping will get a phase $\phi$ enclosed
by the central third triangle of the next-nearest-neighbor. The random
value $\phi_{i}$ is added on the top of the initial flux configurations.
$\phi_{i}$ is uniformly distributed within $[-U_{d}/2,U_{d}/2]$
with $U_{d}$ the random flux strength. (b) Phase diagram of the Haldane
model as a function of $\phi$ and $M/(3\sqrt{2})t_{2}$. To study
the random flux effect, we focus on the fixed value $\phi=-\frac{\pi}{3}$
and take three representative points $A,B$, and $C$. (c) The Bott
index as a function of random flux strength $U_{d}$ corresponding
to points $A,B$, and $C$ in (b). (d) Bott index as a function of $U_d$ at $m=0,\phi=-0.1\pi$. Other parameters are: $t_{1}=1$
and $t_{2}=0.1$. Here we take $100$ random configurations for all
plots. \label{fig:Haldanemodel}}
\end{figure}

Historically, the Haldane model on the honeycomb
lattice was the first model to realize Chern insulator using a staggered
flux configuration \citep{Haldane88prl}. The model
is sketched in Fig. \ref{fig:Haldanemodel}(a).
A magnetic flux of value $-\phi$ is enclosed in the central third
of the triangle of the next nearest-neighbor hopping band. In the
central hexagon (shadowed region), there is a magnetic flux of value
$+6\phi$ such that the whole region contains net flux zero. Without
random flux, the next nearest-neighbor hopping will get a phase $\phi$
in the clockwise direction and a phase $-\phi$ in the counterclockwise
direction {[}see Fig. \ref{fig:Haldanemodel}(a){]}.
The staggered potentials are added by value $M$ on sublattices $A$
and $-M$ on sublattices $B$, respectively, to open the band gap
by breaking the chiral symmetry. The tight-binding Hamiltonian is
\begin{equation}
H_{h}=\sum_{i}M_{i}c_{i}^{\dagger}c_{i}+\left[t_{1}\sum_{\langle i,j\rangle}c_{i}^{\dagger}c_{j}+t_{2}\sum_{\langle\langle i,j\rangle\rangle}e^{i\phi v_{ij}}c_{i}^{\dagger}c_{j}+h.c.\right],
\end{equation}
where $t_{1}$ ($t_{2}$) is the nearest-neighbor (next nearest-neighbor)
hopping energy and $\langle\langle i,j\rangle\rangle$ represents
next nearest-neighbor sites. Here $v_{ij}$ takes 1 or -1 according
to the phase rules as stated above. In the limit $M=0$ and $t_{2}=0,$
the model is just graphene with gap closed at Dirac points. If $M$
is nonzero while keep $t_{2}$ being zero, the system trivially opens
gap at all Dirac points. The interesting case happens if $M$ and
$t_{2}$ are both nonzero: varying these two parameters can make the
system gap close and then reopen, which may indicate a topological
phase transition. Thus Chern number can take nonzero values, for example
$C=1$ in the case of $M-3\sqrt{3}t_{2}\sin\phi>0$ and $M+3\sqrt{3}t_{2}\sin\phi<0$.
The phase diagram of Haldane model in the clean limit is plotted as
in Fig.\ref{fig:Haldanemodel}(b). 

We add random flux in Haldane model on the top of
original flux configurations. The each flux $-\phi$ in the hexagon
now plusses a random value $\phi_{i}$. Here, The random value $\phi_{i}$
is uniformly distributed within $[-U_{d}/2,U_{d}/2]$ with $U_{d}$
the random flux strength. The new flux in a hexagon is still kept
as zero. To show the random-flux-induced phase transition, we focus
on the fixed value of $\phi$ at $\phi=-\frac{\pi}{3}$ and add the
random flux. The parameter $t_{2}$ is set at $t_{2}=0.1$. We tune
the value of $M$ such that the system can be a Chern insulator {[}points
A and B in Fig. \ref{fig:Haldanemodel}(b){]}
or normal insulator (point C). We employ the Bott index as in the
main text to show the topological phase transitions driven by random
flux. In the calculations, the position operators need to be redefined
on the honeycomb lattice.

As shown in Fig. \ref{fig:Haldanemodel}(c),
the Chern insulators persist random flux at weak strength. It is interesting
to see that the Chern insulator survives although the random flux
has heavily driven the magnetic flux enclosed by the nearest-neighbor
hopping from the original value $\phi$ to a random value $\phi+\phi_{i}$.
This result could relax the strict condition for the experimental
realization of Haldane model since the magnetic flux does not necessarily
need to be fixed exactly at the original value of $\phi$. As the
$U_{d}$ increases to a critical point, the Bott index drops suddenly
from $B=1$ to $B=0$, indicating a topological phase transition from
Chern insulators to normal insulators. Comparing point A and B in
Fig.\ref{fig:Haldanemodel}(c), it is easier
to have phase transition with a smaller band gap. If we start with
normal insulator {[}point C in Fig. \ref{fig:Haldanemodel}(c){]},
there is no phase transition as increasing $U_{d}$. Note that even for the case $m=0$, the phase transition happens at finite $U_d$ [Fig.\ref{fig:Haldanemodel}(d)]. Different from
the anisotropic model in the main text, there is also no weak topological
insulators in the isotropic Haldane model. It will be interesting
to add anisotropy by changing the three nearest-neighbor hoppings
to $t_{a}\neq t_{b}\neq t_{c}$ and study the influence of random
flux on phase transitions. 

\bibliographystyle{apsrev4-1-etal-title}

\end{document}